\documentclass[aps,pra,twocolumn,amsmath,amssymb,nofootinbib,superscriptaddress]{revtex4-1}

\usepackage{times}
\usepackage[pdftex]{graphicx}
\usepackage{dcolumn}
\usepackage{bm}
\usepackage{amsmath}
\usepackage{indentfirst}
\usepackage{float}
\usepackage[colorlinks]{hyperref}
\usepackage[dvipsnames]{xcolor}

\usepackage{xcolor}

\begin{document}

\title{Demonstration of Topological Data Analysis on a Quantum Processor}

\author{He-Liang Huang}
\affiliation{Hefei National Laboratory for Physical Sciences at Microscale and Department of Modern Physics,\\
University of Science and Technology of China, Hefei, Anhui 230026, China}
\affiliation{CAS Centre for Excellence and Synergetic Innovation Centre in Quantum Information and Quantum Physics,\\
University of Science and Technology of China, Hefei, Anhui 230026, China}
\affiliation{CAS-Alibaba Quantum Computing Laboratory, Shanghai 201315, China}
\affiliation{Henan Key Laboratory of Quantum Information and Cryptography, Zhengzhou, Henan 450000, China}
\author{Xi-Lin Wang}
\affiliation{Hefei National Laboratory for Physical Sciences at Microscale and Department of Modern Physics,\\
University of Science and Technology of China, Hefei, Anhui 230026, China}
\affiliation{CAS Centre for Excellence and Synergetic Innovation Centre in Quantum Information and Quantum Physics,\\
University of Science and Technology of China, Hefei, Anhui 230026, China}
\affiliation{CAS-Alibaba Quantum Computing Laboratory, Shanghai 201315, China}

\author{Peter P. Rohde}
\affiliation{Centre for Quantum Software $\&$ Information (QSI), Faculty of Engineering $\&$ Information Technology, University of Technology Sydney, NSW 2007, Australia}

\author{Yi-Han Luo}
\author{You-Wei Zhao}
\author{Chang Liu}
\author{Li Li}
\author{Nai-Le Liu}
\author{Chao-Yang Lu}
\author{Jian-Wei Pan}
\affiliation{Hefei National Laboratory for Physical Sciences at Microscale and Department of Modern Physics,\\
University of Science and Technology of China, Hefei, Anhui 230026, China}
\affiliation{CAS Centre for Excellence and Synergetic Innovation Centre in Quantum Information and Quantum Physics,\\
University of Science and Technology of China, Hefei, Anhui 230026, China}
\affiliation{CAS-Alibaba Quantum Computing Laboratory, Shanghai 201315, China}

\date{\today}


\begin{abstract}
Topological data analysis offers a robust way to extract useful information from noisy, unstructured data by identifying its underlying structure. Recently, an efficient quantum algorithm was proposed [Lloyd, Garnerone, Zanardi, Nat. Commun. \textbf{7}, 10138 (2016)] for calculating Betti numbers of data points -- topological features that count the number of topological holes of various dimensions in a scatterplot. Here, we implement a proof-of-principle demonstration of this quantum algorithm by employing a six-photon quantum processor to successfully analyze the topological features of Betti numbers of a network including three data points, providing new insights into data analysis in the era of quantum computing.
\end{abstract}

\maketitle
In exploratory data analysis and data mining, our data often encodes extremely valuable information, but is typically large, unstructured, noisy, and incomplete, such that extracting useful information from the data is an important yet challenging task. Topological data analysis (TDA) \cite{carlsson2009topology} provides a general framework for studying such data in a manner that is insensitive to the particular metric  and robust against noise. In particular, persistent homology \cite{edelsbrunner2002topological, zomorodian2005computingAA} has been well established as a technique for extracting useful information by identifying topological features of data. One essential feature is the number of ${k}$-dimensional holes and voids in datasets, that is, the ${k}$-th Betti number $\beta _k$ (a topological invariant). For instance, the first three Betti numbers, $\beta _0$, $\beta _1$ and $\beta _2$, represent respectively the number of connected components, one-dimensional holes, and two-dimensional voids. The Betti numbers abstract away the actual data, reducing it to a purely topological representation, which is valuable for understanding the underlying structure of datasets. The field of using topological data analysis to analyze Betti numbers of data has been growing rapidly in recent years, yielding applications in image recognition \cite {carlsson2008local}, signal processing \cite{perea2015sliding}, network science \cite{petri2013networks, petri2013topological}, sensor analysis \cite{de2007homological, de2007coverage, de2004topological, ghrist2005coverage}, brain connectomics \cite{giusti2016two, giusti2015clique}, and fMRI data analysis \cite{petri2014homological, lord2016insights}, just to name a few.

Practically however, when facing the issue of computational complexity, classical topological methods pose a formidable task: a set of ${n}$ data points possesses ${{2^n}}$ potential subsets that could contribute to the topology, quickly overwhelming even the most powerful classical computers, even for not-so-large datasets. So far the best classical algorithm for estimating Betti numbers to all orders with accuracy ${\delta }$ takes time ${O({2^n}\mathrm{log}(1/\delta ))}$ \cite{cohen2007stability, basu1999bounding, basu2003different, basu2008computing, basu2014algorithms, friedman1998computing}. Moreover, exact calculation of Betti numbers is known to be PSPACE-hard for some classes of topologies \cite{scheiblechner2007complexity}.

Recently, Lloyd ${et}$ ${al.}$ \cite{lloyd2014topological, lloyd2016quantum} extended methods from quantum machine learning to TDA for efficiently estimating Betti numbers to all orders.  Indeed, if the proportion of $k$-simplices generated from a dataset is large enough, the quantum algorithm for calculating Betti numbers to all orders with accuracy $\delta$ has runtime  ${O({n^5}/\delta )}$ -- exponentially faster than the best known classical algorithms. Furthermore, the algorithm does not require a large-scale quantum random access memory (qRAM)  \cite{giovannetti2008quantum} -- just ${O({n^2})}$ bits is sufficient for the algorithm to store the information of all pairwise distances between the ${n}$ data points. The potential computational speedup and its practicality will likely make quantum TDA a promising application for future quantum computers, in addition to Shor's algorithm \cite{shor1999polynomial, lu2007demonstration, PhysRevLett.99.250505, huang2017experimental}, quantum simulation \cite{feynman1982simulating, lloyd1996universal, lu2009demonstrating, lanyon2010towards}, solving linear systems \cite{harrow2009quantum, cai2013experimental}, and classification of linear vectors \cite{PhysRevLett.113.130503, lloyd2014quantum, cai2015entanglement}.

Here we report a proof-of-principle demonstration of the quantum TDA algorithm on a small-scale photonic quantum processor for the first time. The topological features of Betti numbers of three data points are revealed and monitored at two different topological scales in our experiment. Our experiment successfully demonstrates the viability of the algorithm and suggests that data analytics may be an important future application for quantum computing, with widespread applications in our increasingly data-centric world.

To calculate Betti numbers, we first represent data topologically in terms of relationships between data points. Using a cutoff distance $\epsilon$, we group data points into \emph{simplices} (see Fig.~\ref{fig:simplex_complex}(a))-- fully-connected subsets of data points. The set of simplices forms a \emph{simplicial complex}, the topological structure from which features such as Betti numbers can be extracted. This topological construction is shown in Fig.~\ref{fig:simplex_complex}(b-d).

By determining the complete set of Betti numbers over the full range of $\epsilon$, we can then construct the \emph{barcode} (see Fig.~\ref{fig:simplex_complex}(e)) \cite{Ghrist2008barcodesAA}, a parameterized version of Betti numbers in a distance-dependent manner. Each bar in the region of $\mathbf{H}_k$ represents a $k$-dimensional hole, and the length of the bar indicates its persistence in the parameter $\epsilon$. With the barcode, we can qualitatively filter out the short bars as topological noise and capture the long bars as significant features, since the length of bars is indicative of their persistence against changes in distance $\epsilon$. In Fig.~\ref{fig:simplex_complex}(e), a bar in the region of $\mathbf{H}_1$ persists for a long range, leading us to determine that the underlying topological feature of the unstructured data (Fig.~\ref{fig:simplex_complex}(b)) is a circle.

In general, the quantum TDA algorithm has two main steps (see Fig.~\ref{fig:circuit}(a)). First, one accesses the data to construct the uniform mixture of the $k$-simplices that encode the desired topological structure. The time of this step is in the worst case exponential and in fact depends on the proportion of $k$-simplices. In cases where this fraction is large enough, this step can be implemented efficiently either classically, or using Grover's algorithm, yielding a further quadratic algorithmic enhancement. In the quantum algorithm, this step could be realized via two small steps, namely: (1a) simplicial complex state preparation; (1b) uniform mixed state construction. Second, one implements step (2) to reveal the topological invariants of the structure. This step is realized using the phase-estimation algorithm \cite{nielsen2010quantum}, which provide an exponential speedup over known classical procedures on a quantum computer, in fact \cite{lloyd2014topological, lloyd2016quantum} showed that this can executed in time ${O({n^5}/\delta )}$, with accuracy ${\delta}$. The steps of the quantum algorithm are now described in more detail.

\begin{figure}[tb]
\includegraphics[width=\columnwidth]{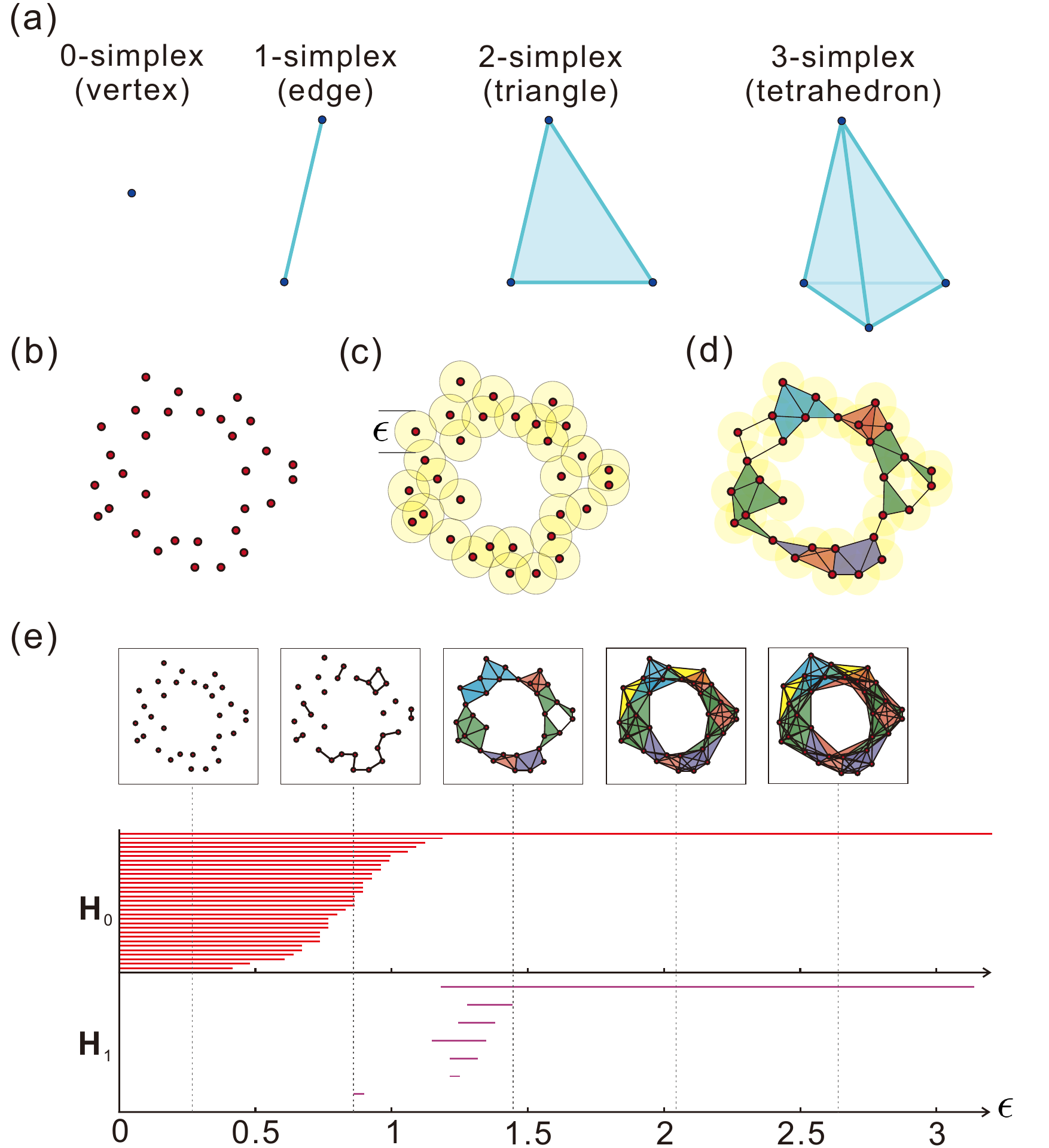}
\caption{(a) $k$-simplices (shown for \mbox{$k=0,1,2,3$}) are fully-connected sets of $k+1$ data points. (b) Scatterplot of data points. (c) Using some arbitrary metric for quantifying \emph{distance} $\epsilon$ between data points, data points within $\epsilon$ of one another receive an edge between them. (d) The simplicial complex is formed as the set of simplices. The colored regions indicate the different simplices within the complex. (e) Construction of the barcode. The horizontal axis represents the distance $\epsilon$. The bars are constructed such that the number of bars that intersect the vertical line through any $\epsilon$ in the area of $\mathbf{H}_k$ equals the Betti number $\beta _k$.} \label{fig:simplex_complex}
\end{figure}

\begin{figure}[tb]
\includegraphics[width=\columnwidth]{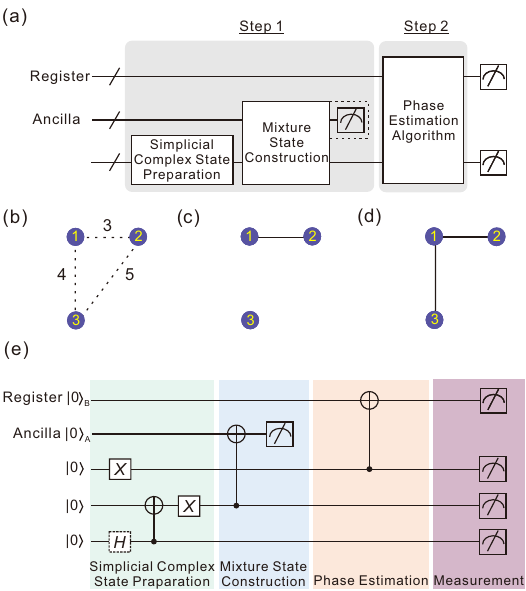}
\caption{Quantum circuit for quantum TDA. (a) Outline of the original quantum circuit. (b) A scatterplot including three data points. (c) Graph representation of the 1-simplices state ${|\varphi \rangle _1^{{\epsilon _1}} = |110\rangle }$ for ${3 < {\epsilon _1} < 4}$. The first and second data points are connected by an edge. (d) Graph representation of 1-simplices state ${|\varphi \rangle _1^{{\epsilon _2}} = (|110\rangle  + |101\rangle )/\sqrt 2}$ for ${4 < {\epsilon _2} < 5}$. The first data point is connected to the second and third points by two edges. (e) Optimized circuit with 5 qubits. The blocks with different colors represent the four basic stages.} \label{fig:circuit}
\end{figure}

Implementing step (1a) constructs the simplicial complex. For a scatterplot including $n$ data points, a $k$-simplex $s_k$ consists of $k+1$ points $V_{j_0},V_{j_1},\dots,V_{j_k}$, together with $k(k + 1)/2$ edges, creating a fully connected subset of the data. We can encode a $k$-simplex as an $n$-qubit quantum state $|s_k\rangle$ with ${k+1}$ 1s at positions ${{j_0},{j_1},...,{j_k}}$ and 0s at the other remaining positions.

The Vietoris-Rips simplicial complex ${S_k^\epsilon }$ is the set of ${k}$-simplices where all points are within distance ${\epsilon }$ of each other. In the quantum implementation, we can construct the simplicial complex state ${|\psi \rangle _k^\epsilon }$ as the uniform superposition of ${k}$-simplices in the complex

\begin{align}
|\psi \rangle _k^\epsilon  = \frac{1}{{\sqrt {|S_k^\epsilon |} }}\sum_{{s_k} \in S_k^\epsilon } {|{s_k}\rangle }.
\end{align}

Classically verify whether all points in each of the $s_k$ are within distance ${\epsilon }$ of each other could help us construct the simplicial complex state. Besides, we can also implement a multi-target Grover's algorithm \cite{grover1997quantum} with a membership oracle function $\{f_k^\epsilon ({s_k}) = 1$ if ${s_k} \in S_k^\epsilon \}$ to verify whether ${{s_k} \in S_k^\epsilon }$, yielding a quadratic speedup. Let ${H_k^\epsilon }$ be the Hilbert space spanned by ${{\rm{|}}{s_k}\rangle }$ where ${{s_k} \in S_k^\epsilon }$. The construction of ${|\psi \rangle _k^\epsilon }$ also reveals the number of ${k}$-simplices, ${|S_k^\epsilon | = \dim H_k^\epsilon }$, and takes time ${O({n^2}{(\zeta _k^\epsilon )^{ - 1/2}})}$, where ${\zeta _k^\epsilon  = |S_k^\epsilon |/ \tbinom {n}{k+1} }$ is the proportion of ${k}$-simplices that are actually in this complex at scale ${\epsilon }$, and ${(\zeta _k^\epsilon)^{-1/2}  = (|S_k^\epsilon |/ \tbinom {n}{k+1}) ^{ - 1/2}}$ is the number of iterations of the multi-target Grover's algorithm. When the proportion is too small, the quantum search procedure will fail to find the simplices \cite{lloyd2014topological, lloyd2016quantum}.

In step (1b), we construct the mixed state,
\begin{align}
{\rho _k^\epsilon  = \frac{1}{|S_k^\epsilon |}\sum_{{s_k} \in S_k^\epsilon } {|{s_k}} \rangle \langle {s_k}{\rm{|}}},
\end{align}
the uniform mixture over the set of simplices in the complex. This procedure can be easily realized by adding an $n$-qubit ancillary register, performing controlled-NOT (CNOT) operations to copy $|\psi \rangle _k^\epsilon$ to construct ${\frac{1}{{\sqrt {|S_k^\epsilon |} }}\sum\limits_{{s_k} \in S_k^\epsilon } {|{s_k}\rangle }  \otimes |{s_k}\rangle }$, and finally tracing out the ancillary register to obtain ${\rho _k^\epsilon }$.

Step (2) acta on the simplicial complex to reveal topological features -- the core of exponential speedup in the algorithm. Define the boundary map ${\partial _k^\epsilon }$ that operates from ${H_k^\epsilon }$ to ${H_{k - 1}^\epsilon }$ by,
\begin{align}
\partial _k^\epsilon |{s_k}\rangle  = \sum\limits_l {{{( - 1)}^l}} |{s_{k - 1}}(l)\rangle,
\end{align}
where ${|{s_{k - 1}}(l)\rangle }$ is obtained from ${{s_k}}$ with vertices ${{j_0}...{j_l}...{j_k}}$ by omitting the ${l}$-th point ${{j_l}}$ from ${{s_k}}$. The ${k}$-th Betti number is defined as \cite{basu1999bounding,basu2003different,basu2008computing,basu2014algorithms},
\begin{align} \label{eq:betti}
{\beta _k^\epsilon  = \dim({\rm{Ker}}\,\partial _k^\epsilon /{\rm{Im}}\,\partial _{k{\rm{ + }}1}^\epsilon )}.
\end{align}

Classical algorithms for calculating Betti numbers to all orders with accuracy $\delta$ require time ${O({2^n}\mathrm{log}(1/\delta ))}$ \cite{cohen2007stability,basu1999bounding,basu2003different,basu2008computing,basu2014algorithms,friedman1998computing}. In quantum TDA, an exponential speedup is achieved by employing the phase-estimation algorithm. For this purpose, the boundary map is embedded into a Hermitian matrix,
\begin{align}
B_k^\epsilon  = \left( {\begin{array}{*{20}{c}}0&{\partial _k^\epsilon }\\{\partial {{_k^\epsilon }^\dag }}&0\end{array}} \right).
\end{align}
Now applying phase-estimation to decompose ${\rho _k^\epsilon }$ in terms of the eigenvectors and eigenvalues of ${B_k^\epsilon }$, one obtains the probability ${\eta _k^\epsilon }$ of projecting onto the kernel by measuring the eigenvalue register. Then the dimension of the kernel of ${\partial _k^\epsilon }$ can be calculated as ${\dim({\rm{Ker}}\,\partial _k^\epsilon) = \eta _k^\epsilon  \cdot |S_k^\epsilon |}$. When both  ${\dim({\rm{Ker}}\,\partial _k^\epsilon })$  and  ${\dim({\rm{Ker}}\,\partial _{k + 1}^\epsilon )}$ are determined, we can reconstruct the ${k}$-th Betti number by,
\begin{align}
\beta _k^\epsilon =& \dim({\rm{Ker}}\,\partial _k^\epsilon)  - \dim({\mathop{\rm Im}\nolimits} \,\partial _{k{\rm{ + }}1}^\epsilon) \nonumber \\
=& \dim({\rm{Ker}}\,\partial _k^\epsilon)  + \dim({\rm{Ker}}\,\partial _{k{\rm{ + }}1}^\epsilon) - |S_{k + 1}^\epsilon |.
\end{align}

\begin{figure*}[htbp]
	\centering
\includegraphics[width=1.5\columnwidth]{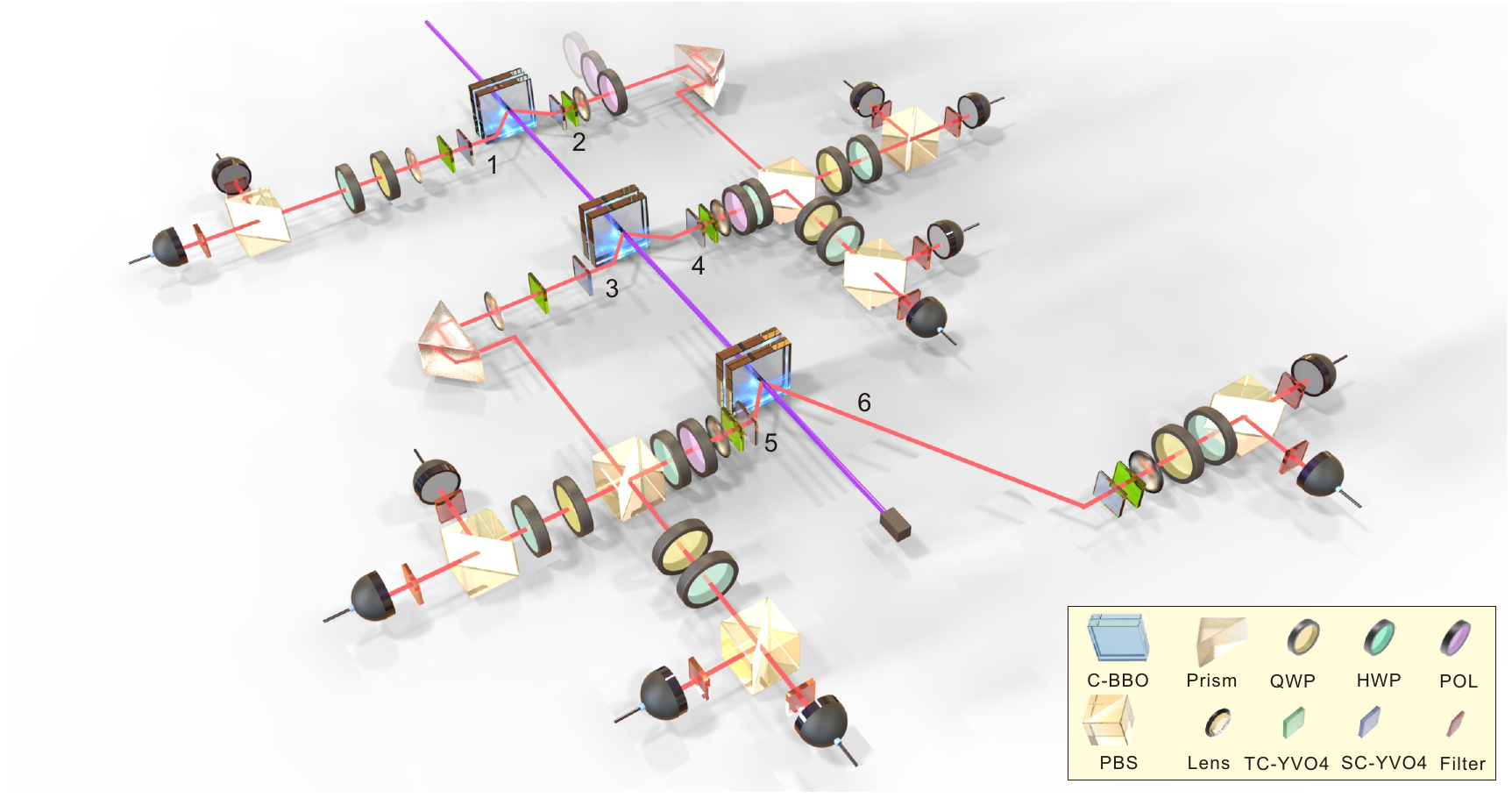}
\caption{Experimental setup. Ultraviolet laser pulses with a central wavelength of 394 nm, pulse duration of 150 fs, and repetition rate of 80 MHz pass through three HWP-sandwiched $\beta$-barium borate (BBO) crystals \cite{wang2016experimental} to produce three entangled photon pairs \mbox{${\left( {|H\rangle |V\rangle  + |V\rangle |H\rangle } \right)/\sqrt 2}$} (see Supplement 1 for details) in spatial modes 1-2, 3-4 and 5-6. Photons 2(3) and 4(5) are temporally and spatially superposed on a PBS. All photons are spectrally filtered with 3-nm bandwidth filters. C-BBO: sandwich-like BBO+HWP+BBO combination; QWP: quarter-wave plate; POL: polarizer; SC-YVO4: YVO4 crystal for spatial compensation; TC-YVO4: YVO4 crystal for temporal compensation.} \label{fig:exp_layout}
\end{figure*}

We note that for some special cases for ${{\kern 1pt} {\kern 1pt} \partial _k^\epsilon }$, it is trivial to calculate ${\dim({\rm{Ker}}\,\partial _k^\epsilon )}$. For example, if a ${k}$-simplex does not exist, ${\dim( {\rm{Ker}}\, \partial _k^\epsilon)  = |S_k^\epsilon | = 0}$, while ${\dim({\rm{Ker}}\, {\partial _0})}$ is always equal to the number of points.

Careful evaluation indicates that step (2) can estimate Betti numbers to all orders with accuracy ${\delta}$ in time ${O({n^5}/\delta )}$ \cite{lloyd2014topological, lloyd2016quantum}. Hence, while in the worst case that their proportion is too small, step (1) will fail to find the $k$-simplices, since both the classical and quantum algorithm will take exponential time. There are specific cases, in particular where step (1) can be implemented efficiently, where the overall quantum algorithm can provide exponential savings. In fact we have tested a particular case using data-points with random distances between them and showed that indeed step (1) can be implemented efficiently (see Supplement 1 for details), either by a classical algorithm or further improving the time by a square root factor through Grover's algorithm.


To experimentally demonstrate the quantum TDA algorithm, we choose the simplest meaningful instance: estimating the Betti numbers for three data points at two different scales. Assume the distances between the three points are 3, 4 and 5 (see Fig.~\ref{fig:circuit}(b)). For scales in the ranges ${3 < {\epsilon _1} < 4}$ and ${4 < {\epsilon _2} < 5}$, the corresponding states for 1-simplices (the ${k}$-simplex for ${k > 1}$ doesn't exist since not all three data points can be connected at $\epsilon _1$ and $\epsilon _2$) are ${|\varphi \rangle _1^{{\epsilon _1}} = |110\rangle }$ (Fig.~\ref{fig:circuit}(c)) and  ${|\varphi \rangle _1^{{\epsilon _2}} = (|110\rangle  + |101\rangle )/\sqrt 2}$ (Fig.~\ref{fig:circuit}(d)) respectively, which means ${|S_1^{{\epsilon _1}}| = 1}$ and ${|S_1^{{\epsilon _2}}| = 2}$. A simple quantum circuit is designed to prepare ${|\varphi \rangle _1^{{\epsilon _1}}}$ (${|\varphi \rangle _1^{{\epsilon _2}}}$) directly by removing (adding) a Hadamard gate marked by dashed lines at step (1) in Fig.~\ref{fig:circuit}(e).

To construct the corresponding uniform mixed states, we don't actually need to generate a complete copy of ${|\varphi \rangle _1^{{\epsilon _1}}}$ (${|\varphi \rangle _1^{{\epsilon _2}}}$). Instead, we need only perform a CNOT operation between the auxiliary qubit ${|0{\rangle _A}}$ and the second qubit of ${|\varphi \rangle _1^{{\epsilon _1}}}$ (${|\varphi \rangle _1^{{\epsilon _2}}}$) to partially copy the state of simplices. After tracing out the ancillary qubit, the uniform mixed states ${\rho _{}^{{\epsilon _1}}}$ and ${\rho _{}^{{\epsilon _2}}}$ are obtained.

Next, apply quantum phase-estimation to reveal information related to Betti numbers. Since there are only three data points, ${k}$-dimensional holes for ${k > 1}$ can not exist. Therefore, only the 0-th and 1-st Betti numbers need to be calculated. We note that the algorithm cares not about the exact eigenvalue spectrum, but the probability of detecting ${|0\rangle }$ in the eigenvalue register. We can exploit this property to reduce the number of qubits required in the eigenvalue register. A particular treatment for boundary matrices is utilized to greatly simplify the complex circuit (see Supplement 1 for details) -- a single CNOT operation between the eigenvalue register comprising only one qubit ${|0{\rangle _B}}$ and the first bit of ${\rho _{}^{{\epsilon _1}}}$ (${\rho _{}^{{\epsilon _2}}}$) is sufficient for realizing phase-estimation. Finally, the information related to Betti numbers will be read out by measuring the eigenvalue register. Note that since the quantum TDA algorithm only depends on how the points are connected, not the precise distances between points, our circuit works for all nontrivial cases of three points (where one, or two edges are present). The cases where zero or three edges are present are trivial, since we could clearly know the Betti numbers in the cases that the $N$ points are all disconnected ($\beta _0=N$, and $\beta _k=0$ for $k>0$) or all connected ($\beta _0=1$, and $\beta _k=0$ for $k>0$) for $N$ points without calculating.

Fig.~\ref{fig:exp_layout} shows the setup of our experiment. We use single photons as qubits, where the logical qubits ${|0\rangle }$ and ${|1\rangle }$ are encoded into horizontal (${H}$) and vertical (${V}$) polarization, respectively. With these settings, the step of simplices state preparation becomes straightforward. ${|\varphi \rangle _1^{{\epsilon _1}} = |V{\rangle _3}|V{\rangle _2}|H{\rangle _1}}$ and ${|\varphi \rangle _1^{{\epsilon _2}} = {(|V\rangle _3}|V{\rangle _2}|H{\rangle _1} + |V{\rangle _3}|H{\rangle _2}|V{\rangle _1})/\sqrt 2}$ can be prepared directly by adding or removing the polarizer in path 2 respectively, where the index ${i}$ in ${|H(\mathrm{or}\, V){\rangle _i}}$ denotes the spatial mode. Photons 4 (ancilla) and 5 (eigenvalue register) are both disentangled by polarizers into ${|H\rangle }$, and then photons 3 and 6 (trigger) immediately collapse into ${|V\rangle }$. Note that the CNOT gates can be simulated using combinations of a polarizing beam splitter (PBS) and a half-wave plate (HWP) \cite{lu2007demonstration}, since the target qubits are fixed at ${|H\rangle }$. This setup, in principle, suffices to demonstrate the underlying conceptual principles of quantum TDA.

Before running the algorithm, we first characterized the performance of the optical quantum circuit. In the case of ${4 < {\epsilon _2} < 5}$, a three-photon entangled state ${|\phi \rangle  = (|H{\rangle _1}|V{\rangle _2}|V{\rangle _4} + |V{\rangle _1}|H{\rangle _2}|H{\rangle _4})/\sqrt 2}$ is generated after implementing the CNOT gate in step (1a). We measured the fidelity of the experimentally prepared state (see Supplement 1 for details) as ${F=0.954(6)}$, which exceeds the threshold of 0.5 for the entanglement witness to confirm genuine multi-partite entanglement \cite{guhne2009entanglement}. To the best of our knowledge, such a high fidelity for three photon entanglement has never been achieved before \cite{hamel2014direct}.

After tracing out the ancilla in the Pauli-$Z$ basis, the uniform mixed states,
\begin{align}
\rho^{{\epsilon _1}} &= |V{\rangle _3}|V{\rangle _2}|H{\rangle _1}\langle V{|_3}\langle V{|_2}\langle H{|_1}, \nonumber \\
\rho^{\epsilon _2} &= (|V\rangle_3 |V\rangle_2 |H\rangle_1 \langle V|_3 \langle V |_2 \langle H |_1 \nonumber \\
&+ |V\rangle_3 |H\rangle_2 |V\rangle_1 \langle V|_3 \langle H|_2\langle V|_1)/2,
\end{align}
are created at the scales of ${3 < {\epsilon _1} < 4}$ and ${4 < {\epsilon _2} < 5}$, respectively. We characterized these states using quantum state tomography to reconstruct the density matrices. (See Supplement 1 for details). The fidelity ${F_p = {(\mathrm{Tr}\sqrt {{\rho ^{1/2}}{\rho _{\exp }}{\rho ^{1/2}}} )^2}}$ and trace distance ${D(\rho ,{\rho _{\exp }}) = \mathrm{Tr}|\rho  - {\rho _{\exp }}|/2}$ between the reconstructed (${{\rho _{\exp }}}$) and ideal (${{\rho}}$) matrices were calculated as ${{F_p^{{\rho ^{{\epsilon _1}}}}} = 0.9817(9)}$, ${{F_p^{{\rho ^{{\epsilon _2}}}}} = 0.9819(10)}$ and ${{D_{{\rho ^{{\epsilon _1}}}}} = 0.0183(5)}$, ${{D_{{\rho ^{{\epsilon _2}}}}} = 0.0181(5)}$ respectively. Furthermore, The fidelity ${F_p}$ and trace distance ${D}$ are related by the inequality ${1 - \sqrt {{F_p}}  \le D \le \sqrt {1 - {F_p}} }$ \cite{nielsen2010quantum}. In our experiment, both ${{D_{{\rho ^{{\epsilon _1}}}}}}$ and ${{D_{{\rho ^{{\epsilon _2}}}}}}$ are located in the range of ${0.009 \le D \le 0.135}$, and close to the lower bound.


The final results were read out via 6-fold coincidence events. Figures.~\ref{fig:results}(a,b) show the measurement results of the eigenvalue register at the scales of ${3 < {\epsilon _1} < 4}$ and ${4 < {\epsilon _2} < 5}$, respectively. In the case of ${3 < {\epsilon _1} < 4}$, with a probability of ${\eta _1^{{\epsilon _1}} = 0.045(14)}$ we measure ${|0\rangle }$ in the eigenvalue register, from which we calculate the dimension of the kernel space as ${\dim({\rm{Ker}}\,\partial _1^{{\epsilon _1}}) = \eta _1^{{\epsilon _1}} \cdot |S_1^{{\epsilon _1}}| = 0.045(14)}$. Since ${\dim({\rm{Ker}}\,\partial _0^\epsilon) = 3}$ and ${\dim({\rm{Ker}}\,\partial _2^\epsilon)  = |S_2^\epsilon | = 0}$ for ${\epsilon  = {\epsilon _1}\,(\mathrm{or}{\kern 1pt} {\kern 1pt} {\epsilon _2})}$, we finally obtain the 0-th Betti number ${\beta _0^{{\epsilon _1}} = 2.045(14)}$ and 1-st Betti number ${\beta _1^{{\epsilon _1}} = 0.045(14)}$, following Eq.~\ref{eq:betti}, which can be rounded to ${\beta _0^{{\epsilon _1}} = 2}$ and ${\beta _1^{{\epsilon _1}} = 0}$. In the case of ${4 < {\epsilon _2} < 5}$, the probability of measuring ${|0\rangle }$ in the eigenvalue register is 0.038(9). Using the same approach, we calculate the 0-th and 1-st Betti numbers as ${\beta _0^{{\epsilon _2}} = 1.076(18)}$ and ${\beta _1^{{\epsilon _2}} = 0.038(9)}$, respectively, which can be rounded to ${\beta _0^{{\epsilon _2}} = 1}$ and ${\beta _1^{{\epsilon _2}} = 0}$. That is to say, we have revealed and tracked the topological features of the dataset in  Fig.~\ref{fig:circuit}(b) at two different scales: the number of connected components at scales of ${{\epsilon _1}}$ and ${{\epsilon _2}}$ are 2 and 1, respectively, and no $k$-dimensional holes for ${k > 1}$ exist. From these results,  the barcode is constructed as shown in Fig.~\ref{fig:results}(c).

\begin{figure}[tb]
\includegraphics[width=\columnwidth]{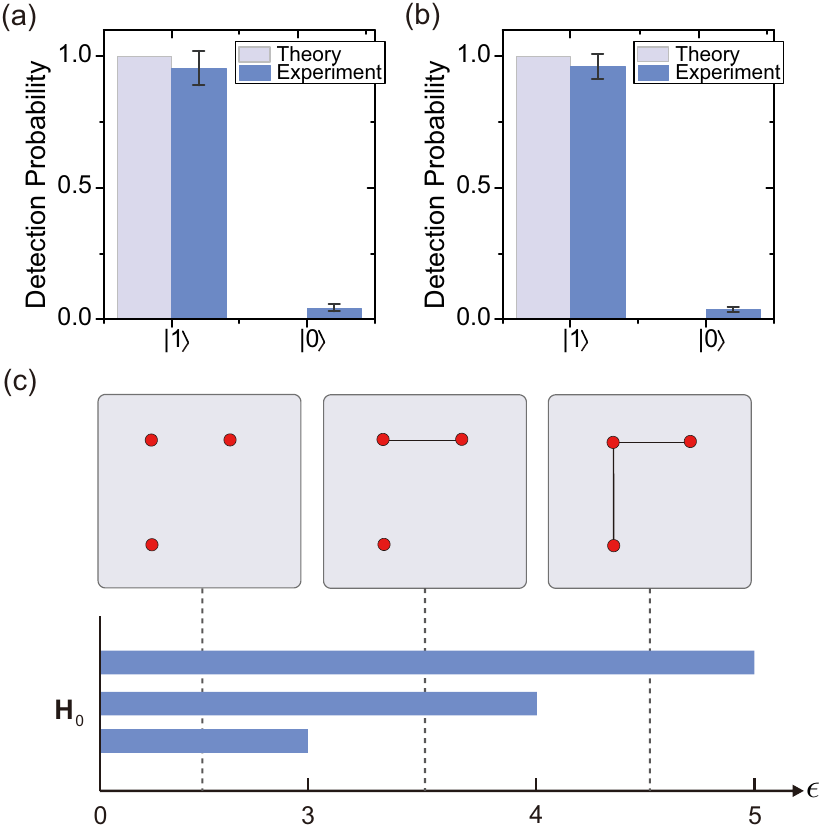}
\caption{Final experimental results. The output is determined by measuring the eigenvalue register in the Pauli-$Z$ basis. Measured expectation values (blue bars) and theoretically predicted values (gray bars) are shown for two different 1-simplices state inputs: (a) ${|\varphi \rangle _1^{{\epsilon _1}} = |110\rangle }$, (b) ${|\varphi \rangle _1^{{\epsilon _2}} = (|110\rangle  + |101\rangle )/\sqrt 2}$. Error bars represent one standard deviation, deduced from propagated Poissonian counting statistics of the raw detection events. (c) The barcode for $0<\epsilon<5$.  Since no $k$-dimensional holes for $k\ge1$ exist at these scales, only the 0-th Betti barcode is given here. For $0<\epsilon<3$, there is no connection between each point, so the 0-th Betti number is equal to the number of points. That is, there are three bars at $0<\epsilon<3$. At scales of ${3<{\epsilon _1}<4}$ and ${4<{\epsilon _2}<5}$ , the 0-th Betti number are 2 and 1.} \label{fig:results}
\end{figure}

To further quantify the experimental performance, we use the similarity measure ${\gamma  = {\left(\sum\nolimits_{k = 0}^1 \sqrt{e_k t_k} \right)^2}}$ \cite{Fuchs1996} to characterize the overlap between experimental and theoretical values, where ${{e_k}}$ and ${{t_k}}$ are the experimental and theoretical output probabilities of the state ${|k\rangle }$, respectively. The data in Fig.~\ref{fig:results} shows the results as ${{\gamma _{{\epsilon _1}}} = 0.955(3)}$ and ${{\gamma _{{\epsilon _2}}} = 0.962(2)}$, indicating near perfect experimental accuracy, confirming that the algorithm is successful.


We note that for the quantum TDA algorithm, the results are read out by measuring the eigenvalues. In general, the eigenvalue register requires only a few qubits for the quantum TDA algorithm (1 qubit in the current work), since we only care about the proportion of ${|0\rangle }$ in the eigenvalue register, rather than the exact value of all eigenvalues. Thus, a small amount of measurements are sufficient for obtaining reliable results, an important feature for the scalability of the algorithm.

In addition, theoretically, for the quantum TDA algorithm, only the qubits in the eigenvalue register need to be measured, rather than having to measure all qubits. In our experiment, since the photons generated by spontaneous parametric down conversion are probabilistic, to ensure that all qubits in the circuit have been generated, and the quantum circuits have been fully implemented, we need to measure 6-fold coincidence events. In fact, this is a common problem encountered in the current linear optical quantum computing. Fortunately, with the development of deterministic quantum dot single photon source \cite{he2017deterministic}, and other techniques \cite{kaneda2015time}, we believe this problem can eventually be overcome. We anticipate that with more qubits (more photons \cite{wang2016experimental,wang2017high} or higher dimensional states \cite{fickler2012quantum, wang2015quantum}), our proposal could be extended to the analysis of much larger datasets in the future.

In summary, we have presented the first proof-of-principle demonstration of quantum TDA on a small-scale photonic quantum processor. The topological features of a dataset comprising three data points is revealed and tracked at two different topological scales, fully reproducing the Betti numbers associated with the topology of the data. Future advances in the field could open up new frontiers in data analysis for quantum computing, including signal and image analysis, astronomy, network and social media analysis, behavioral dynamics, biophysics, oncology and neuroscience.

\textbf{Acknowledgements:} We thank R.-Z. Liu, Michele Cirafici, T. L. for enlightening discussions. This work was supported by the National Natural Science Foundation of China, the Chinese Academy of Sciences, and the National Fundamental Research Program. P.P.R. is funded by an ARC Future Fellowship (project FT160100397).
\newline
\newline
\noindent See Supplement 1 for supporting content.

\rule{0.42\textwidth}{0.5mm}
\maketitle
\section*{Supplemental Material}

\section{Background and practical applications of Betti number and TDA}

\begin{figure}[h]
\includegraphics[width=\columnwidth]{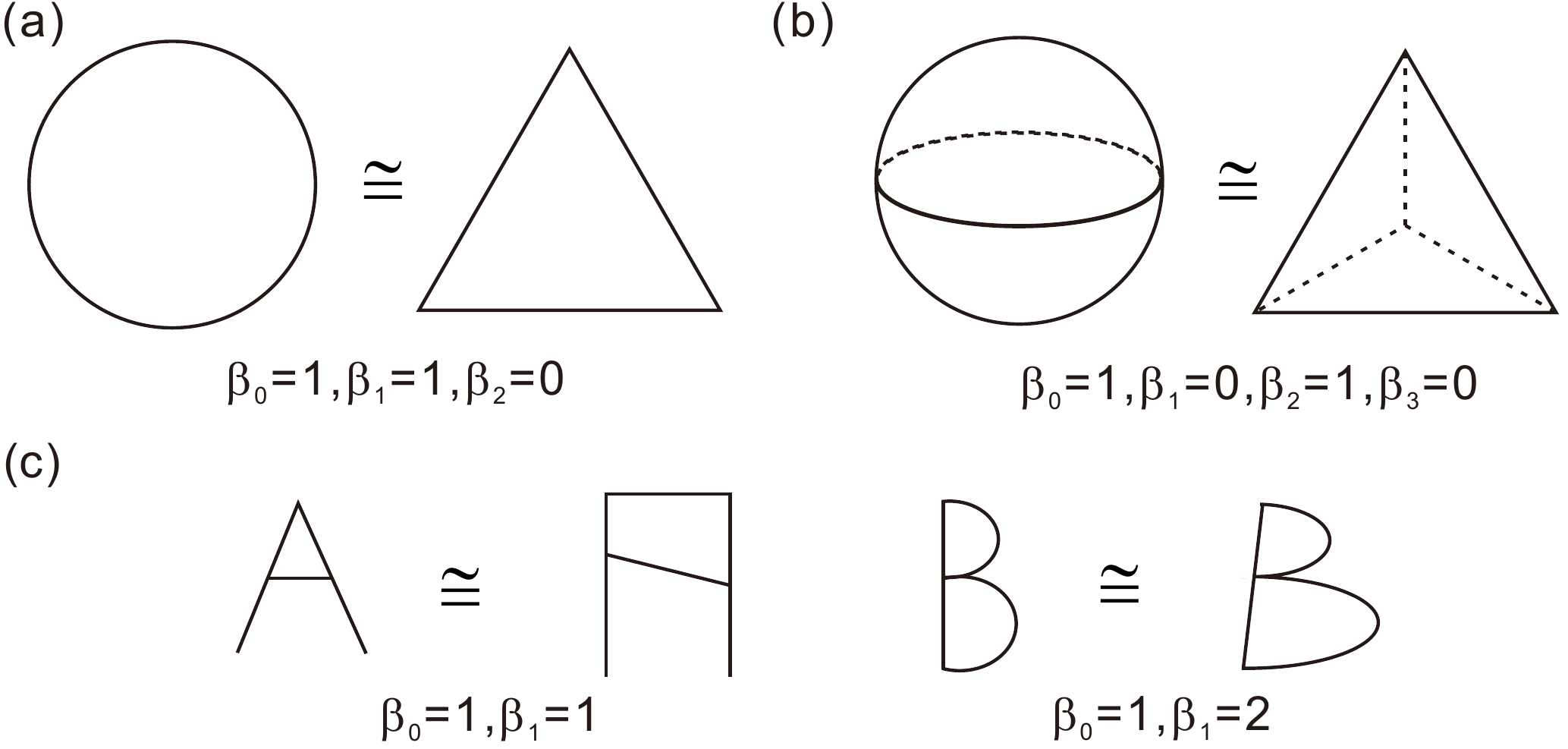}
\caption{Several examples for the explanation of Betti numbers, demonstrating their ability to capture structural information even in the presence of local deformations.} \label{fig:betti}
\end{figure}

Betti numbers are a way to describe the connectivity within a topological space. In simplest terms, the $k$-th Betti number $\beta_k$ counts the the number of $k$-dimensional holes in a topological space, for example,

- $\beta_0$ is the number of connected components;

- $\beta_1$ is the number of planar holes (1-dimensional holes);

- $\beta_2$ is the number of  two-dimensional voids (2-dimensional holes);

- ...

Betti numbers are topological invariants. If two Betti numbers are the same for two different spaces then the spaces are homotopy equivalent \cite{carlsson2009topology}. To demonstrate Betti numbers more vividly, some examples are shown in Fig.~\ref{fig:betti}. We can see that a circle has a connected component, a 1-dimensional holes, thus $\beta_0=1,\beta_1=1$. The Betti numbers of circle are the same as a triangle, so they are are homotopy equivalent (see Fig.~\ref{fig:betti}(a)); Similarly, the two-dimensional hollow sphere is homotopy equivalent to a hollow tetrahedron (see Fig.~\ref{fig:betti}(b)). Thus, Betti numbers can record significant topological features of a shape, which could be directly used in pattern recognition \cite{carlsson2014topological}, anomaly detection \cite{johannsen2012betti}, computational linguistics \cite{nilsson2013topology}. For instance, considering a simple shape recognition task, namely the recognition of printed letters, by using the Betti numbers, we could identify and distinguish the letters ``A'' and ``B'' in Fig.~\ref{fig:betti}(c), even in the presence of some deformation.

Now, we briefly introduce some mathematical background for Betti numbers. For more details, one can refer to \cite{nakahara2003geometry}.

\rule{0.42\textwidth}{0.25mm}

We first describe how to use a simplicial complex to formally describe a topological structure.

\emph{Simplex}: A $k$-simplex $\sigma_k=[V_{j_0}, \cdots, V_{j_k}]$ is a fully connected set of $k+1$ affine geometric points $V_{j_0}, \cdots, V_{j_k}$, together with $k(k + 1)/2$ edges (see Fig 1(a) for some example). where $k$ is the dimension of the simplex.

\emph{Simplicial complex}: Roughly speaking, a simplicial complex $K$ is a finite set simplices (see Fig. 1(d) for an example) such that:

$i$) any face of a simplex of $K$ is a simplex of $K$,

$ii$) the intersection of any two simplices of $K$ is either empty or a common face of both.

Next, we will introduce the chain group, boundary operator, cycle group and boundary group, and then how to calculate the Betti numbers.

$k$-\emph{chain group:} A $k$-chain is a formal sum of $k$-simplices with integer coefficients, which can be written as $c = \sum\limits_{i = 1}^p {{\varepsilon _i}} {\sigma _i}$ with ${\varepsilon _i} \in {{\rm Z}_2}$, where $\left\{ {{\sigma _1}, \cdots ,{\sigma _p}} \right\}$  is the set of $k$-simplices of $K$. The set of all $k$-chains forms an Abelian group $C_k(K)$.

$k$-\emph{boundary operator:} For a $k$-simplex $\sigma_k=[V_{j_0}, \cdots, V_{j_k}]$, the boundary map ${\partial _k}:{C_k}(K) \to {C_{k - 1}}(K)$ is given by
\begin{center}
${\partial _k}(\sigma ) = \sum\limits_{i = 0}^k {{{( - 1)}^i}[{V_{{j_0}}}, \cdots ,{{\hat V}_{{j_i}}}, \cdots ,{V_{{j_k}}}]}$
\end{center}

where ${{\hat V}_{{j_i}}}$ indicates that ${{V}_{{j_i}}}$ is removed, and ${[{V_{{j_0}}}, \cdots ,{{\hat V}_{{j_i}}}, \cdots ,{V_{{j_k}}}]}$ is the $k-1$-simplex spanned by all the vertices except ${{V}_{{j_i}}}$.

$k$-\emph{boundary group} and $k$-\emph{cycle group}: The $k$-\emph{boundary group} is defined as ${B_k}(K) = \mathrm{Im}~\partial_{k+1}=\{ c \in {C_k}(K)|\exists c' \in {C_{k + 1}}(K),{\partial _{k + 1}}(c') = c\}$, containing elements that are boundaries of $k+1$-dimensional objects; The $k$-\emph{cycle group} is defined as $Z_k(K)=\mathrm{Ker}~\partial_k=\{c\in C_k(K)|\partial_k c=0\}$, the elements in the cycle group can be understood as `loops'. It can be proved that ${B_k}(K) \subseteq {Z_k}(K) \subseteq {C_k}(K)$.

\emph{Homology group}: Let $K$ be an $k$-dimensional simplicial complex. The $k$th homology group $H_k(K)$ associated with $K$ is defined by $H_k(K)\equiv Z_k(K)/B_k(K)$, which represents those elements of $Z_k(K)$ (loops) that are not boundaries.

\emph{Betti numbers}: The $k$-th Betti number $\beta_k$ is defined by $\beta_k(K)\equiv \mathrm{dim}~H_k(K)=\dim({\rm{Ker}}\,\partial _k /{\rm{Im}}\,\partial _{k{\rm{ + }}1}).$

\rule{0.42\textwidth}{0.25mm}

Using Betti numbers, we can detect invisible geometric features of high-dimensional objects. Applying Betti numbers to data analysis could help us analyze and exploit the complex topological and geometric structures underlying data. Next, we will introduce how to use persist homology, a  sophisticated topological data analysis method, to extract useful information by identifying the topological features (Betti numbers) of data.

\emph{From points to simplicial complex}: In data analysis, data is usually represented as an unordered sequence of points (see Fig. 1(b)), to analyze the Betti numbers of data, requiring a method to construct a simplicial complex.

To define a simplicial complex, the most obvious way is to use the points as the vertices of a combinatorial graph whose edges are determined by proximity. Using a cutoff distance $\epsilon$, and connecting points within distance $\epsilon$ (see Fig 1.~(b-d) for the procedure), we can construct the simplicial complex (see Fig 1.~(d)), called a Vietoris-Rips simplicial complex.

\emph{Computing Betti numbers}: Having constructed the simplicial complex of data points, we use the method above to calculate Betti numbers, finding the topological structure of the data points.

\emph{Barcode}: Converting data points into a simplicial complex requires a choice of parameter -- cutoff distance $\epsilon$. However, if $\epsilon$ is too small, almost all points are separated, and no overall structure is apparent; if $\epsilon$ is too large, all the points may be connected with each other, the complex is a single high dimensional simplex, and no topological holes exist. It is challenging to select an appropriate scale for a given dataset. To address this problem, we observe the evolution of topological features for the full range of $\epsilon$, rather than focussing on a particular numeric value, yielding the barcode (see Fig.~1(e)). Each bar in the region of $\mathbf{H}_k$ of the barcode represents a $k$-dimensional hole, the length of which indicates its persistence in the parameter $\epsilon$. With the barcode, we can qualitatively filter out the short bars as topological noise and capture the long bars as significant, persistent topological features, since the length of bars is indicative of their persistence against changes in distance $\epsilon$. For further details, refer to \cite{zomorodian2005computing}.

There are many interesting and useful applications of topological data analysis. For instance, in the field of image recognition, Carlsson et al. found that high-contrast 3$\times$3 pixel patches from grayscale digital images concentrate near the surface of a Klein bottle in a higher-dimensional space \cite{carlsson2008local}; in the field of signal processing, Perea and Harer found that persistent homology can detect periodicity in time-series data preventing noise \cite{perea2015sliding}, which is very stable and accurate especially in the presence of damping; in unsupervised machine learning, persistent homology also provides a powerful tool for the analysis of musical data, exploring common features of classical scores \cite{sethares2014topology}.

\section{Numerical simulation of the proportion of $k$-simplices in some cases}
\begin{figure}[htbp]
\includegraphics[width=0.9\columnwidth]{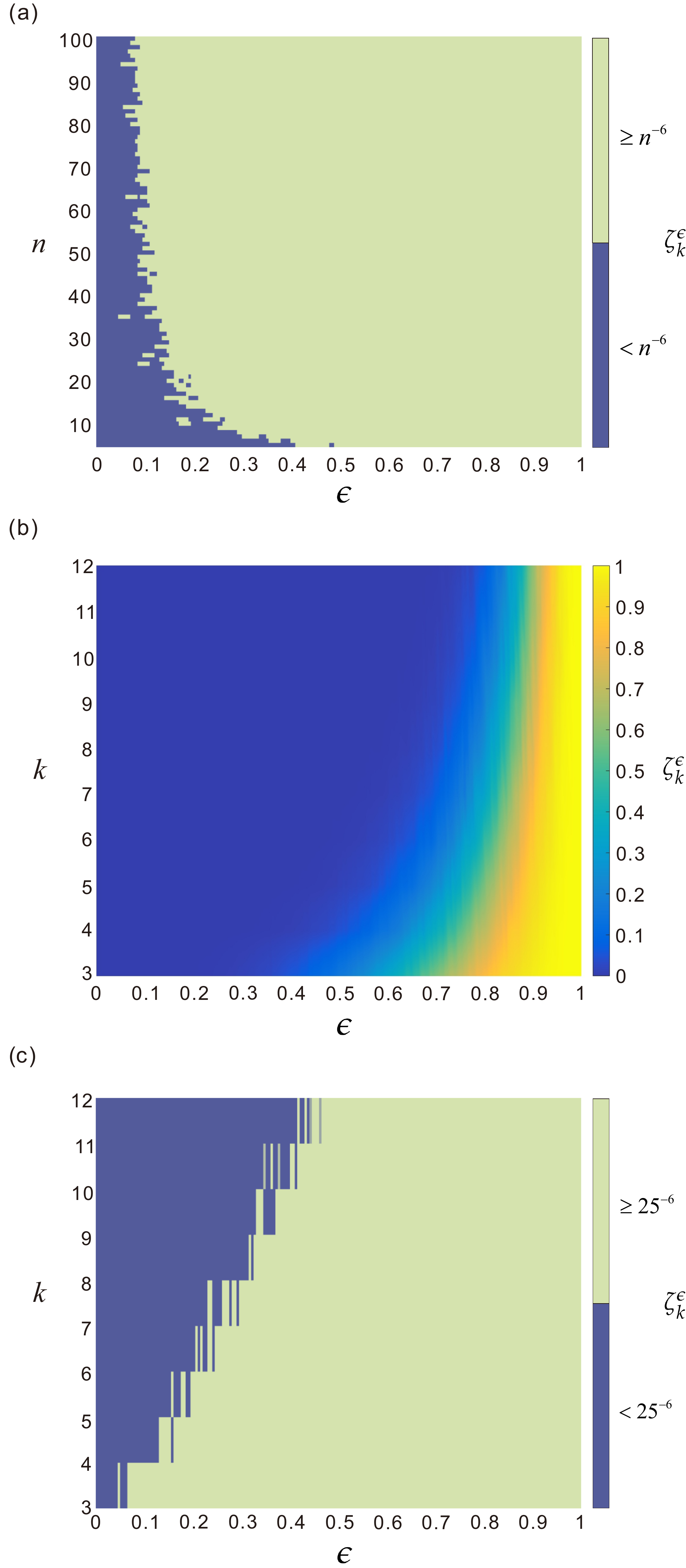}
\caption{The proportion of $k$-simplices in our numerical simulations. (a) Let $k=4$, the relationship among the proportion of ${k}$-simplices $\zeta _k^\epsilon$, the number of data point $n$ ($y$ axis) and cutoff distance $\epsilon$ ($x$ axis). The blue area represents $\zeta _k^\epsilon < n^{-6}$, the green area represents $\zeta _k^\epsilon \ge n^{-6}$. (b) Let $n=25$, the relationship among the proportion of ${k}$-simplices, the dimension $k$ of the ${k}$-simplices and cutoff distance $\epsilon$. (c) Let the blue area represent $\zeta _k^\epsilon < n^{-6}$ in (b), and the green area represent $\zeta _k^\epsilon \ge n^{-6}$ in (b). It is clear that the green area is far larger than the blue area.} \label{fig:ratio}
\end{figure}

As mentioned in the main text, the efficiency of step (1) depends on the proportion of ${k}$-simplices. Here, we studied the relationship among the proportion of ${k}$-simplices, the number of data point $n$, the dimension $k$ of the ${k}$-simplices, and cutoff distance $\epsilon$ by numerical simulation (see Fig.~\ref{fig:ratio}).

In our simulations, without loss of generality, we randomly set the distances between different points in the range of [0,1]. In Fig.~\ref{fig:ratio}(a), we take $k=4$ as an example to simulate the relationship among the proportion of ${k}$-simplices, the number of data points $n$ and cutoff distance $\epsilon$. Since the computational complexity of step (1) in quantum TDA is ${O({n^2}{(\zeta _k^\epsilon )^{ - 1/2}})}$, and the computational complexity of step (2) is ${O({n^5}/\delta )}$, where ${\delta}$ is the accuracy, we could regard step (1) as efficient in quantum TDA if ${{n^2}{(\zeta _k^\epsilon )^{ - 1/2}}} \le {n^5}/\delta $, that is $\zeta _k^\epsilon \ge n^{-6}$. In Fig.~\ref{fig:ratio}(a), the blue area represents $\zeta _k^\epsilon < n^{-6}$, and the green area represents $\zeta _k^\epsilon \ge n^{-6}$. We can see that, as $n$ increases, the the green area becomes larger and the blue area becomes smaller. Thus, with the increase of $n$, the step (1) is efficient at a wider range of cutoff distance $\epsilon$.

In Fig.~\ref{fig:ratio}(b), we take $n=25$ as an example to simulate the relationship between the proportion of ${k}$-simplices, their dimension $k$, and the cutoff distance $\epsilon$. It is clear that the proportion of ${k}$-simplices becomes smaller gradually at each cutoff distance $\epsilon$ as $k$ becomes larger. Similar to Fig.~\ref{fig:ratio}(a), we let the blue area represent $\zeta _k^\epsilon < n^{-6}$, and the green area represent $\zeta _k^\epsilon \ge n^{-6}$, yielding Fig.~\ref{fig:ratio}(c). We can see that even when $k=12$ and $\tbinom {n}{k+1}$ reaches the maximum $\tbinom {25}{12}$, the green area can still encompass over 50\% of the region. Obviously, by analyzing all three figures in Fig.~\ref{fig:ratio}, the regime of step (1) that can be regarded as efficient is much larger than than that regarded as inefficient. That is, step (1) can be implemented efficiently in the cases of our numerical simulations.

\section{Experimental Errors analysis}
In this section, we will analyze errors introduced by experimental noise and provide an error threshold analysis.

The imperfections in our experiment can be attributed to two major causes: higher-order photon emissions, and partial distinguishability of independent photons. In order to suppress the influence of higher-order photon emissions, we placed two single-photon detectors at each measurement port. This dual-channel setup can partially suppress higher-order events where both detectors trigger simultaneously at one measurement port, indicating the presence of multiple photons. To ensure the high levels of indistinguishability between independent photons, all photons are spectrally filtered by 3-nm narrow-band filters.

\begin{figure}[htbp]
\includegraphics[width=\columnwidth]{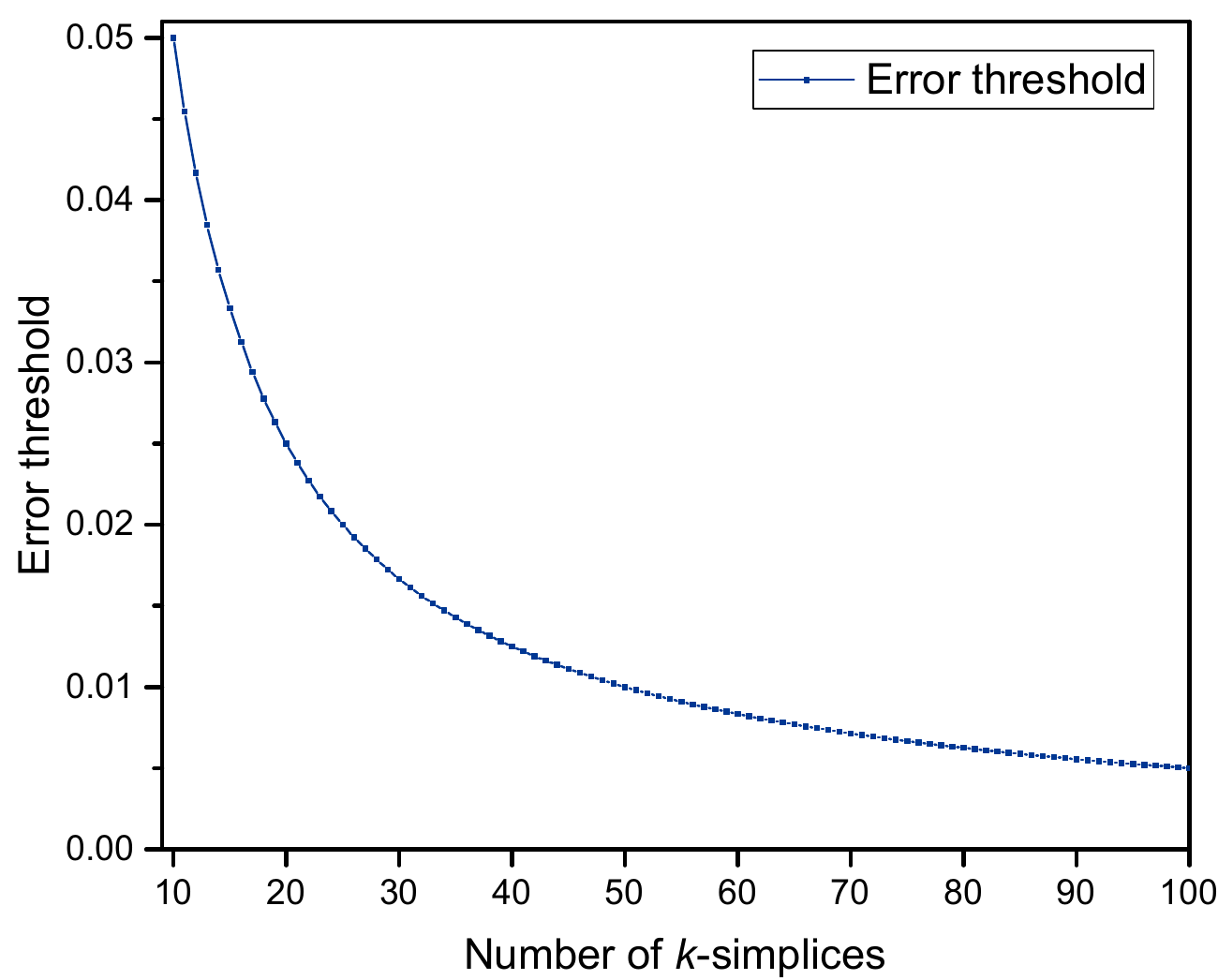}
\caption{The relationship between the number of $k$-simplices $|{S_k}|$ ($x$ axis) and error threshold $E_t$ ($y$ axis). Obviously, as $|{S_k}|$ increases, the error threshold $E_t$ decreases.} \label{fig:error}
\end{figure}

The final result of the quantum TDA algorithm is decided by the probability of the zero eigenvalue measured in the eigenvalue register. Assume the ideal probability of measuring the zero eigenvalue is ${\eta _i}$, then the dimension of the kernel of $\partial _k^\epsilon$ could be calculated as $\dim (Ker\partial _k^\epsilon ) = {\eta _i} \cdot |{S_k}|$. To obtain the correct dimension in the experiment, we need to ensure that $|\dim (Ker\partial _k^\epsilon )_{ideal}-\dim (Ker\partial _k^\epsilon )_{experiment}|< 0.5$, that is $|{\eta _e} - {\eta _i}| \cdot |{S_k}| < 0.5$ if we use the rounding principle, where ${\eta _e}$ is the probability of the experimentally measured zero eigenvalue. To quantify the experimental error threshold, we define the error as $E_t=|{\eta _e} - {\eta _i}|$, and then simulate the error threshold that satisfies the constraint condition $|{\eta _e} - {\eta _i}| \cdot |{S_k}| < 0.5$. The relationship between the number of $k$-simplices $|{S_k}|$ ($x$ axis) and error threshold ($y$ axis) is shown in Fig.~\ref{fig:error}. Obviously, as $|{S_k}|$ increases, the error threshold decreases. Thus, appropriate fault-tolerance mechanisms should be employed when we deal with large-scale dataset.

Note that unlike the the previous quantum algorithm, the quantum TDA algorithm only cares about the probability of the zero eigenvalue, not  all the individual values in the eigenvalue register. Thus, the quantum TDA algorithm, in principle, could be more robust to noise than other algorithms, such as Shor's algorithm \cite{shor1999polynomial} and the HHL algorithm\cite{harrow2009quantum}, which require an exact quantum state as output.

\section{necessity of constructing the mixed state}

In the quantum TDA algorithm, step (1) is used to construct the uniform mixture of the $k$-simplices, which is realized by: (1a) simplicial complex state preparation; (1b) uniform mixed state construction. In fact, the purpose of step (1) is to sample a uniform $k$-simplex, which is the essential reason for constructing mixed state.

Next, we will provide the reason why the quantum TDA algorithm can not directly use the pure state generated in step (1a) as the input of step (2). In step (2), we use quantum phase-estimation algorithm to decompose a mixed state in terms of the eigenvectors of the Hermitian matrix $B_k$, which acts on the space $H_{k - 1}^\epsilon \oplus H_k^\epsilon$, and find the probability of the zero eigenvalue to compute the dimension of the kernel of $\partial _k^\epsilon$. The mixed state is
$$
\rho^\epsilon_k=\frac{1}{|S^\epsilon_k|}\sum_{s_k\in S^\epsilon_k}|s_k\rangle\langle s_k|.
$$
where each $k$-simplices $|s_k\rangle$ is the basis, and $\rho^\epsilon_k$ is a maximally mixed state. According to quantum mechanics, even using another complete basis set, the maximally mixed state $\rho^\epsilon_k$ is still of the above form. Thus, $\rho^\epsilon_k$ could be rewritteb as the eigenstate set $\{|n_k\rangle\}$ of $\partial _k^\epsilon$
$$
\rho^\epsilon_k=\frac{1}{|S^\epsilon_k|}\sum_{i=1}^{|S^\epsilon_k|}|n_i\rangle\langle n_i|.
$$

Introduce qubits ${\rm{|}}0{\rangle ^t}$ as the eigenvalue register, after the phase-estimation algorithm,
$${\rm{|}}0{\rangle ^t}\rho _k^\epsilon \xrightarrow{\mathrm{phase-estimation}} \frac{1}{{|S_k^\epsilon |}}\sum\limits_{i = 1}^{|S_k^\epsilon |} | {\lambda _i}\rangle |{n_i}\rangle \langle {n_i}|\langle {\lambda _i}|.$$

For each eigenstate $|n_i\rangle$, the eigenvalue register will output its corresponding eigenvalue ${\rm{|}}{\lambda _i}\rangle$. Thus, The probability of measuring the zero eigenvalue in the register is $N^\epsilon_k(0)/|S^\epsilon_k|$, where $N^\epsilon_k(0)$ is the number of eigenstates in $\{|n_k\rangle\}$ whose eigenvalue is zero, that is, the dimension of the kernel of $\partial _k^\epsilon$. However, if we directly used the pure state generated in step (1a) as the input to step (2), after we decompose the pure state in terms of the eigenvectors of the Hermitian matrix $B_k$, the probability of the zero eigenvalue in the register will be meaningless due to interference effects. For ease of understanding, we will give an example to show that using the pure state as the input of step (2) will output wrong results.

\begin{figure}[htbp]
\center
\includegraphics[width=0.7\columnwidth]{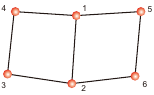}
\caption{A topological structure of six points, the points are connected by 7 edges.} \label{fig:mixeg}
\end{figure}

For the topological structure in Fig.~\ref{fig:mixeg}, the 1-simplices are $|110000\rangle, |011000\rangle, |001100\rangle, |100100\rangle, |100010\rangle, |000011\rangle,$ $|010001\rangle $, which are denoted as $|a\rangle, |b\rangle, |c\rangle, |d\rangle, |e\rangle, |f\rangle, |g\rangle$ respectively. The 0-simplices are $|100000\rangle, |010000\rangle, $ $|001000\rangle, |000100\rangle, $ $ |000010\rangle, |000001\rangle$, which are denoted as $|1\rangle, |2\rangle, |3\rangle, |4\rangle, |5\rangle, |6\rangle$ respectively. The Hermitian operator $B _1$ is

\begin{equation}
B_1=
\left(
\begin{array}{cc}
O & \partial_1 \\
\partial_1^\dagger & O
\end{array}
\right).
\end{equation}

where
\begin{equation}
\partial_1=
\bordermatrix{
&\scalebox{0.8}{$|a\rangle$}&\scalebox{0.8}{$|b\rangle$}&\scalebox{0.8}{$|c\rangle$}&\scalebox{0.8}{$|d\rangle$}&\scalebox{0.8}{$|e\rangle$}&\scalebox{0.8}{$|f\rangle$}&\scalebox{0.8}{$|g\rangle$}\cr
\scalebox{0.8}{$|1\rangle$}&1&0&0&-1&1&0&0\cr
\scalebox{0.8}{$|2\rangle$}&-1&1&0&0&0&0&-1\cr
\scalebox{0.8}{$|3\rangle$}&0&-1&1&0&0&0&0\cr
\scalebox{0.8}{$|4\rangle$}&0&0&-1&1&0&0&0\cr
\scalebox{0.8}{$|5\rangle$}&0&0&0&0&-1&1&0\cr
\scalebox{0.8}{$|6\rangle$}&0&0&0&0&0&-1&1\cr
},
\end{equation}

There are only two eigenstates of the Hermitian matrix $B_1$ whose eigenvalue is zero:

$
|n_1\rangle=1/2~(|a\rangle-|e\rangle-|f\rangle-|g\rangle),
$
$$
|n_2\rangle=1/\sqrt{60}~(3|a\rangle+4|b\rangle+4|c\rangle+4|d\rangle+|e\rangle+|f\rangle+|g\rangle).
$$

Therefore, after the phase-estimation algorithm, the probability of measuring the eigenvalue of zero in eigenvalue register should be 2/7. However, if we use the the pure state,
\[
\begin{split}
|\psi\rangle&=1/\sqrt{7}~(|a\rangle+|b\rangle+|c\rangle+|d\rangle+|e\rangle+|f\rangle+|g\rangle)\\
&\xrightarrow{\mathrm{phase-estimation}}-\frac{1}{\sqrt{7}}|n_1\rangle|0\rangle+\frac{18}{\sqrt{420}}|n_2\rangle|0\rangle+\cdots,
\end{split}
\]
Obviously, the probability of measuring the eigenvalue of zero is ${\left( { - \frac{1}{{\sqrt 7 }}} \right)^2} + {\left( { \frac{{18}}{{\sqrt {420} }}} \right)^2} = \frac{{32}}{{35}}$, which is inconsistent with the expectation 2/7. By this counterexample, we can see that the algorithm can not use pure state generated in step (1a) as the input to step (2).

\section{Circuit details}

To implement the algorithm with a limited number of qubits, our designed circuit differs from the original algorithm via several modifications, some of which have already been mentioned in the main text. Here we show the details of the modifications to phase-estimation, the core of the quantum TDA algorithm. Before introducing the modification, we provide two preliminaries:

(i) Let $U$ be an arbitrary unitary operator, the eigenvector and eigenvalue sets of which are ${\{ {|{u_1}\rangle ,|{u_2}\rangle ,...,|{u_n}\rangle\}}}$ and ${\{ {{\lambda _1},{\lambda _2},...,{\lambda _n}\}}}$, respectively. If we transform the unitary operator $U$ into ${\alpha {U^2}}$, where ${\alpha  \ne 0}$ is a constant, then the eigenvalue set of ${\alpha {U^2}}$ become ${\{ {\alpha {\lambda _1}^2,\alpha {\lambda _2}^2,...,\alpha {\lambda _n}^2}}\}$, and the eigenvector set will not change. We note that if ${{\lambda _i} \ne 0}$, then ${\alpha {\lambda _i}^2 \ne 0}$, else if ${{\lambda _i} = 0}$, then ${\alpha {\lambda _i}^2 = 0}$.

(ii) Suppose ${|0{\rangle ^{ \otimes t}}|u\rangle}$ is the input of the phase-estimation algorithm, where ${|0{\rangle ^{ \otimes t}}}$ is an eigenvalue register with $t$ qubits, and ${|u\rangle}$ is an eigenvector of unitary operator $U$ with eigenvalue ${{e^{2\pi i\phi }}}$ (${\phi  \approx 0.{\phi _1}...{\phi _t}}$ with binary representation). The phase-estimation algorithm is designed to output ${|{\phi _1}...{\phi _t}\rangle |u\rangle }$, where ${ |{\phi _1}...{\phi _t}\rangle}$ is an approximation to the phase ${\phi}$ with a precision of $t$ bits.

Specifically, the Hermitian boundary matrices at scales ${3 < {\epsilon _1} < 4}$ and ${4 < {\epsilon _2} < 5}$ are

\begin{equation}
B^{{\epsilon _1}} =
\bordermatrix{
&\scalebox{0.6}{$|100\rangle$}&\scalebox{0.6}{$|010\rangle$}&\scalebox{0.6}{$|001\rangle$}&\scalebox{0.6}{$|110\rangle$}\cr
\scalebox{0.6}{$|100\rangle$}&0&0&0&-1\cr
\scalebox{0.6}{$|010\rangle$}&0&0&0&1\cr
\scalebox{0.6}{$|001\rangle$}&0&0&0&0\cr
\scalebox{0.6}{$|110\rangle$}&-1&1&0&0\cr
}
\end{equation}
\begin{equation}
B^{{\epsilon _2}} =
\bordermatrix{
&\scalebox{0.6}{$|100\rangle$}&\scalebox{0.6}{$|010\rangle$}&\scalebox{0.6}{$|001\rangle$}&\scalebox{0.6}{$|110\rangle$}&\scalebox{0.6}{$|101\rangle$}\cr
\scalebox{0.6}{$|100\rangle$}&0&0&0&-1&-1\cr
\scalebox{0.6}{$|010\rangle$}&0&0&0&1&0\cr
\scalebox{0.6}{$|001\rangle$}&0&0&0&0&1\cr
\scalebox{0.6}{$|110\rangle$}&-1&1&0&0&0\cr
\scalebox{0.6}{$|101\rangle$}&-1&0&1&0&0\cr
}
\end{equation}

The eigenvalue and eigenvector sets of the boundary matrices ${{B^{{\epsilon _1}}}}$ are ${{\rm{\{ }}\lambda _1^{{\epsilon _1}},\lambda _2^{{\epsilon _1}},\lambda _3^{{\epsilon _1}},\lambda _4^{{\epsilon _1}}{\rm{\}  = \{  - }}\sqrt 2 {\rm{,}}\sqrt 2 {\rm{,0}},{\rm{0\} }}}$ and ${{\rm{\{ |}}\beta _1^{{\epsilon _1}}\rangle ,|\beta _2^{{\epsilon _1}}\rangle ,|\beta _3^{{\epsilon _1}}\rangle ,|\beta _4^{{\epsilon _1}}\rangle {\rm{\} }}}$, respectively, are

\begin{align}
|\beta _k^{{\epsilon _1}}\rangle  = \left\{ {\begin{array}{*{20}{r}}
{\frac{1}{2}|100\rangle  - \frac{1}{2}|010\rangle +\frac{1}{\sqrt 2} |110\rangle ,\,\,k = 1}\\
{ - \frac{1}{2}|100\rangle  + \frac{1}{2}|010\rangle + \frac{1}{\sqrt 2} |110\rangle ,\,\,k = 2}\\
{|001\rangle ,\,\,k = 3}\\
{\frac{1}{\sqrt 2} |100\rangle + \frac{1}{\sqrt 2} |010\rangle ,\,\,k = 4}
\end{array}} \right.
\end{align}

To reduce the number of qubits required in the eigenvalue register, we set ${{B_1} = {({B^{{\epsilon _1}}})^2}/2}$, then the eigenvalue spectrum becomes ${{\rm{\{ }}\lambda _1^{{\epsilon _1}},\lambda _2^{{\epsilon _1}},\lambda _3^{{\epsilon _1}},\lambda _4^{{\epsilon _1}}{\rm{\}  = \{ }}1{\rm{,}}1{\rm{,0}},{\rm{0\} }}}$, without changing the eigenvector set. We note that the algorithm cares not about the full spectrum but the probability of ${|0\rangle}$ being detected in the  register, so this special treatment is justified. Then transforming ${{B_1}}$ into the unitary operator ${{e^{i\pi {B_1}}}}$ allows us to implement phase-estimation using an eigenvalue register with only one qubit ${{\rm{|}}0{\rangle _B}}$. For the input ${|0{\rangle _B}|\varphi \rangle _1^{{\epsilon _1}}}$, we apply the transformation,

\begin{align}
|0{\rangle _B}|\varphi \rangle _1^{{\epsilon _1}} = |0{\rangle _B}|110\rangle  &= \frac{1}{\sqrt 2} |0{\rangle _B}(|\beta _1^{{\epsilon _1}}\rangle  + |\beta _2^{{\epsilon _1}}\rangle ) \nonumber \\
\xrightarrow{\mathrm{phase\,estimation}} &= \frac{1}{\sqrt 2} |1{\rangle _B}(|\beta _1^{{\epsilon _1}}\rangle  + |\beta _2^{{\epsilon _1}}\rangle ) \nonumber \\
&= |1{\rangle _B}|110\rangle = |1{\rangle _B} |\varphi \rangle _1^{\epsilon _1}.
\end{align}

Similarly, at the scale of ${{\epsilon _2}}$, we set ${{B_2} = {({B^{{\epsilon _2}}})^2}}$ and transform ${{B_2}}$ into the unitary operator ${{e^{i\pi {B_2}}}}$ to meet experimental requirements. For the input ${|0\rangle \langle 0{|_B} \otimes \rho _{}^{{\epsilon _2}}}$, the phase-estimation procedure outputs the state ${|1\rangle \langle 1{|_B} \otimes \rho _{}^{{\epsilon _2}}}$, where ${{\rho ^{{\epsilon _2}}} = (|110\rangle \langle 110| + |101\rangle \langle 101|)/2}$.  Thus, in our experiment, only a single CNOT operation between the eigenvalue register comprising only one qubit ${|0{\rangle _B}}$ and the first bit of ${\rho _{}^{{\epsilon _1}}}$ (${\rho _{}^{{\epsilon _2}}}$) is sufficient for us to compile the phase-estimation algorithm.

\section{experimental implementation of the Circuit}
In the experiment, we use single photons as qubits, where the logical qubits ${|0\rangle }$ and ${|1\rangle }$ are encoded into horizontal (${H}$) and vertical (${V}$) polarization, respectively. The setup of our experiment is shown in Fig.~3. Photons in paths 1, 2, and 3 are used to construct simplex states. Photons 4 (ancilla) and 5 (eigenvalue register) are both disentangled by polarizers into ${|H\rangle }$, and then photons 3 and 6 (trigger) immediately collapse into ${|V\rangle }$. Here we describe details of how to experimentally implement the circuit in Fig.~2(b).

\begin{figure}[h]
\includegraphics[width=\columnwidth]{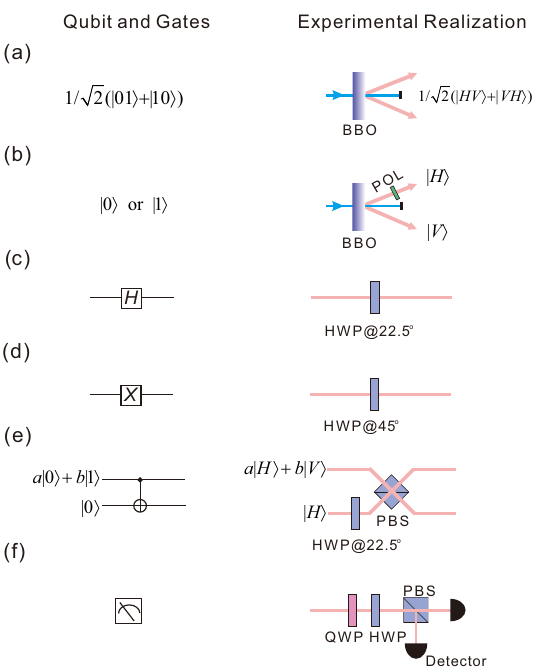}
\caption{The experimental implementation of the circuit. (a) Preparation of the entangled state ${\rm{1/}}\sqrt 2 ({\rm{|}}0\rangle {\rm{|}}1\rangle {\rm{ + |}}1\rangle {\rm{|}}0\rangle )$; (b) Preparation of the quantum state ${\rm{|}}0\rangle$ or ${\rm{|}}1\rangle$; (c) Realization of the $H$ gate; (d) Realization of the $X$ gate; (e) Realization of the CNOT gate; (f) Measurement setup. BBO: $\beta$-barium borate, POL: polarizer, HWP: half-wave plate, QWP: quarter-wave plate, PBS: polarizing beam splitter.} \label{fig:exp-circuit}
\end{figure}

In the initialization stage, the photons in our experiment are generated by spontaneous parametric down-conversion using $\beta$-barium borate (BBO). Ultraviolet laser pulses pass through a BBO crystal to produce entangled state ${\rm{1/}}\sqrt 2 ({\rm{|}}0\rangle {\rm{|}}1\rangle {\rm{ + |}}1\rangle {\rm{|}}0\rangle )$ (see Fig.~\ref{fig:exp-circuit}(a)). If we do not want the entangled state, we could use a polarizer (POL) to disentangle the entangled state to ${\rm{|}}0\rangle$ or ${\rm{|}}1\rangle$ (see Fig.~\ref{fig:exp-circuit}(b)).

In the quantum gate operation stage, we need to implement a $H$ gate, $X$ gate, and CNOT gate. The single-qubit quantum gates $H$ and $X$ can  beexperimentally realized using half-wave plates (HWP) of $22.5^\circ$ (see Fig.~\ref{fig:exp-circuit}(c)) and $45^\circ$ (see Fig.~\ref{fig:exp-circuit}(d)), respectively. Since the target qubit of the CNOT gate in our circuit is $|0\rangle$, it can be realized using a combination of a polarizing beam splitter (PBS) and a HWP, and post-selecting the events where there is exactly one photon exiting each output of the PBS \cite{lu2007demonstration} (see Fig.~\ref{fig:exp-circuit}(e)).

In the measurement stage, each photon passes through a quarter-wave plate (QWP), a HWP, a PBS, and is finally read out by using a single-photon detector (see Fig.~\ref{fig:exp-circuit}(f)). By adjusting the angle of the QWP and HWP, we can measure the photonic qubit in arbitrary bases.

\begin{figure}[ht]
\center
\includegraphics[width=0.9\columnwidth]{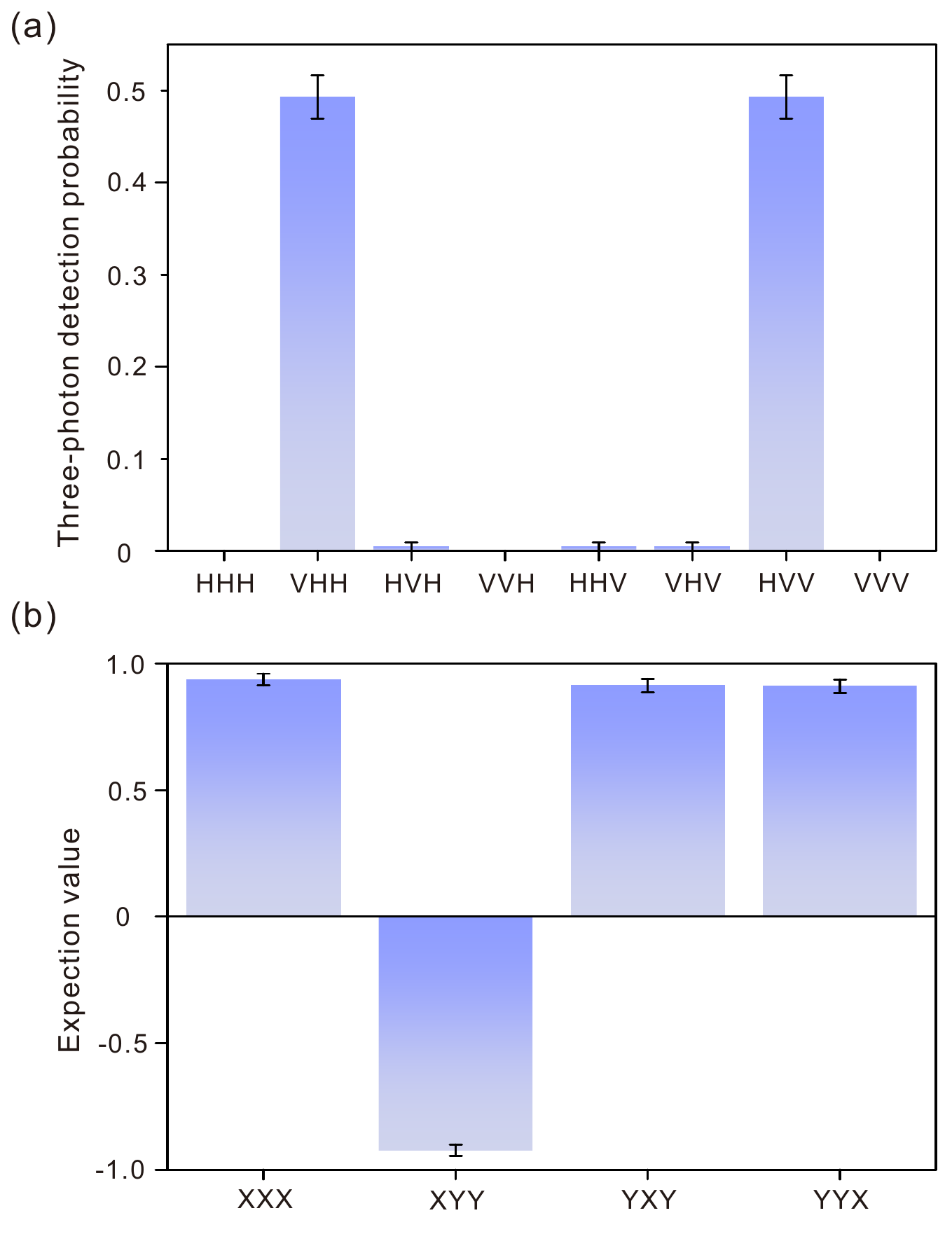}
\caption{Experimental results for entanglement witness measurements. (a) Threefold coincidence detection probabilities in the $H/V$ basis. (b) Expectation values of $XXX$, $XYY$, $YXY$ and $YYX$. Error bars represent one standard deviation, deduced from propagated Poissonian counting statistics of the raw detection events.} \label{fig:S1}
\end{figure}

\section{Photon source}

We developed a high-performance source of polarization entangled photons generated via spontaneous parametric down-conversion (SPDC) using a sandwich-like bulk \cite{wang2016experimental}, which consists of two identically cut 2mm-thick beam-like type-II $\beta$-barium borate (BBO) crystals with one half-wave plate (HWP) inserted between them. The source simultaneously exhibits high brightness (${\sim}$850Hz/mW), high efficiency (${\sim}$45\% collection efficiency with 3nm bandwidth filters, and ${\sim}$88\% collection efficiency without narrowband filtering) and high fidelity (${\sim}$0.98) at a pump power of 240mW. These three essential features are crucial for future scalable photonic quantum technologies.

\section{Characterizing the three-photon entangled state}

Here we show the details for determining the fidelity of the three-photon entangled state ${|\phi \rangle  = (|HVV\rangle  + |VHH\rangle)/\sqrt{2}}$ and verifying genuine multipartite entanglement \cite{seevinck2001sufficient} using an entanglement witness. The fidelity is the overlap of the experimentally produced state ${{\rho _{\exp }}}$ with the desired state ${{\rho _\mathrm{ideal}}}$,
\begin{align}
{F_{|\phi \rangle }} = \langle \phi |{\rho _{\exp }}|\phi \rangle
\end{align}
For the three-photon entangled state ${{\rho _\mathrm{ideal}} = |\phi \rangle \langle \phi | = (|HVV\rangle \langle HVV| + |VHH\rangle \langle VHH|}$ ${+ (XXX + YXY - XYY + YYX)/4)/2}$ where $X$, $Y$ and $Z$ are the Pauli matrices ${{\sigma _x}}$, ${{\sigma _y}}$, ${{\sigma _z}}$ respectively. Fig.~\ref{fig:S1} shows the experimental data. The expectation values of ${|HVV\rangle \langle HVV| + |VHH\rangle \langle VHH|}$ and ${(XXX + YXY - XYY + YYX)/4}$ are 0.987(1) and 0.921(12) respectively. Thus, the state fidelity of ${|\phi \rangle}$ can be calculated as ${{F_{|\phi \rangle }} = 0.954(6)}$, which exceeds the threshold of 0.5 required for the entanglement witness. With high statistical significance (${\sim}$76 standard deviations), genuine three-photon entanglement is confirmed.

\begin{figure} [ht]
\includegraphics[width=\columnwidth]{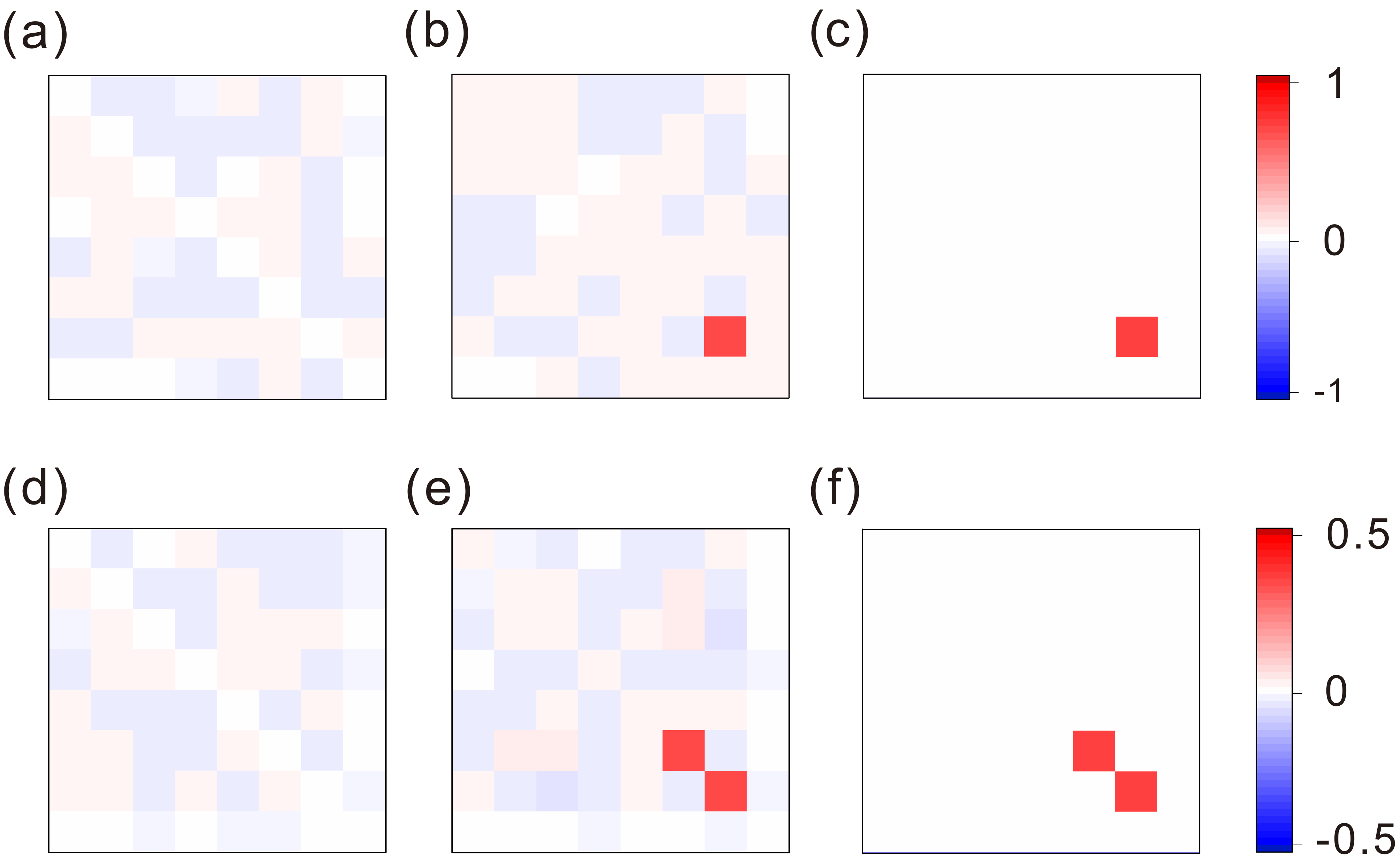}
\caption{Reconstructed and ideal density matrices. (a) Imaginary component of the reconstructed density matrix ${\rho _{\exp }^{{\epsilon _1}}}$. (b) Real component of ${\rho _{\exp }^{{\epsilon _1}}}$. (c) Real part of the theoretically ideal matrix ${\rho _{}^{{\epsilon _1}}}$ (imaginary components are identically zero). (d) Imaginary component of the reconstructed density matrix ${\rho _{\exp }^{{\epsilon _2}}}$. (e) Real component of ${\rho _{\exp }^{{\epsilon _2}}}$. (f) Real part of the theoretically ideal matrix ${\rho _{}^{{\epsilon _2}}}$ (imaginary components are identically zero).} \label{fig:matrices}
\end{figure}

\section{State reconstructions}

The matrix form of the reconstructed experimentally obtained states ${\rho _{\exp }^{{\epsilon _1}}}$ and ${\rho _{\exp }^{{\epsilon _2}}}$ are,

\begin{tiny}
\begin{widetext}
\begin{align}
\rho _{\exp }^{{\epsilon _1}} &= \left( {\begin{array}{*{20}{c}}
{2.78 \times {{10}^{ - 17}}}&0&{2.17 \times {{10}^{ - 19}}i}&0&{0.0046 + 0.0022i}&{0.0012 + 0.0024i}&{0.0029 - 0.0024i}&{ - 0.0003 + 0.0018i}\\
0&0&{ - 8.67 \times {{10}^{ - 19}}i}&0&{0.0012 - 0.0024i}&0&{ - 0.0111 + 0.0053i}&0\\
{2.17 \times {{10}^{ - 19}}i}&{ - 8.67 \times {{10}^{ - 19}}i}&{2.78 \times {{10}^{ - 17}}}&0&{0.0100 + 0.0024i}&{0.0066 + 0.0017i}&{ - 0.0046 - 0.0153i}&{ - 0.0126 + 0.0109i}\\
0&0&0&0&{ - 0.0066 + 0.0006i}&0&{0.0174 - 0.0109i}&0\\
{0.0046 - 0.0022i}&{0.0012 + 0.0024i}&{0.0100 - 0.0024i}&{ - 0.0066 - 0.0006i}&{0.0137}&{0.0020 + 0.0044i}&{ - 0.0047 - 0.0047i}&{0.0513 - 0.0021i}\\
{0.0012 - 0.0024i}&0&{0.0066 - 0.0017i}&0&{0.0020 - 0.0044i}&0&{ - 0.0487 - 0.0271i}&{0.0002i}\\
{0.0029 + 0.0024i}&{ - 0.0111 - 0.0053i}&{ - 0.0046 + 0.0153i}&{0.0174 + 0.0109i}&{ - 0.0047 + 0.0047i}&{ - 0.0487 + 0.0271i}&{0.9863}&{0.0245 + 0.0000i}\\
{ - 0.0003 - 0.0018i}&0&{ - 0.0126 - 0.0109i}&0&{0.0513 + 0.0021i}&{0.0000 - 0.0002i}&{0.0245}&0
\end{array}} \right) \nonumber \\
\rho _{\exp }^{{\epsilon _2}} &= \left( {\begin{array}{*{20}{c}}
{ - 0.0010}&0&{0.0071}&{ - 0.0012}&{0.0020 + 0.0023i}&{ - 0.0014 - 0.0016i}&{0.0050 - 0.0051i}&{ - 0.0025 - 0.0051i}\\
0&{0.0010}&{ - 0.0012}&{ - 0.0024}&{ - 0.0108 - 0.0125i}&{0.0449 - 0.0023i}&{ - 0.0030 - 0.0093i}&{0.0082 - 0.0164i}\\
{0.0071}&{ - 0.0012}&{ - 0.0010}&0&{0.0050 + 0.0329i}&{0.0258 + 0.0010i}&{ - 0.0265 + 0.0023i}&{0.0130 + 0.0133i}\\
{ - 0.0012}&{ - 0.0024}&0&{0.0010}&{ - 0.0037 - 0.0129i}&{ - 0.0107 + 0.0239i}&{ - 0.0106 - 0.0086i}&{ - 0.0102 - 0.0116i}\\
{0.0020 - 0.0023i}&{ - 0.0108 + 0.0125i}&{0.0050 - 0.0329i}&{ - 0.0037 + 0.0129i}&{0.0071}&{0.0044 - 0.0210i}&{0.0071 + 0.0244i}&{0.0087 + 0.0277i}\\
{ - 0.0014 + 0.0016i}&{0.0449 + 0.0023i}&{0.0258 - 0.0010i}&{ - 0.0107 - 0.0239i}&{0.0044 + 0.0210i}&{0.4990}&{ - 0.0181 - 0.0009i}&{0.0165 + 0.0317i}\\
{0.0050 + 0.0051i}&{ - 0.0030 + 0.0093i}&{ - 0.0265 - 0.0023i}&{ - 0.0106 + 0.0086i}&{0.0071 - 0.0244i}&{ - 0.0181 + 0.0009i}&{0.4929}&{ - 0.0197 - 0.0070i}\\
{ - 0.0025 + 0.0051i}&{0.0082 + 0.0164i}&{0.0130 - 0.0133i}&{ - 0.0102 + 0.0116i}&{0.0087 - 0.0277i}&{0.0165 - 0.0317i}&{ - 0.0197 + 0.0070i}&{0.0010}
\end{array}} \right)
\end{align}
\end{widetext}
\end{tiny}

However, the eigenvalue spectra of ${\rho _{\exp }^{{\epsilon _1}}}$ and ${\rho _{\exp }^{{\epsilon _2}}}$  are $\{0.9909, 0.0608, -0.0521, 0.0093, -0.0077, -0.0014, 0.0003, \\-0.0001\}$ and $\{0.5257, 0.4806, 0.05729, -0.0537, -0.0184, \\0.0149, -0.0064, 0.0001\}$ respectively, which violates the positivity of density matrices. To avoid this problem, we employ maximum likelihood estimation \cite{james64measurement} to reconstruct ${\rho _{\exp }^{{\epsilon _1}}}$ and ${\rho _{\exp }^{{\epsilon _2}}}$, obtaining the corresponding legitimate density matrices

\begin{tiny}
\begin{widetext}
\begin{align}
\rho _{\exp }^{{\epsilon _1}} &= \left( {\begin{array}{*{20}{c}}
{0.0015}&{0.0002 - 0.0009i}&{0.0001 - 0.0003i}&{ - 0.0008}&{ - 0.0002 + 0.0002i}&{ - 0.0005 - 0.0005i}&{0.0027 + 0.0024i}&0\\
{0.0002 + 0.0009i}&{0.0019}&{0.0003 - 0.0001i}&{ - 0.0004 - 0.0007i}&{ - 0.0004 - 0.0002i}&{0.0004 - 0.0009i}&{ - 0.0062 + 0.0068i}&0\\
{0.0001 + 0.0003i}&{0.0003 + 0.0001i}&{0.0018}&{0.0001 - 0.0014i}&{0.0003 + 0.0001i}&{0.0006 + 0.0007i}&{ - 0.0109 - 0.0198i}&{0.0003}\\
{ - 0.0008}&{ - 0.0004 + 0.0007i}&{0.0001 + 0.0014i}&{0.0023}&{0.0004 + 0.0005i}&{ - 0.0008 + 0.0010i}&{0.0219 - 0.0114i}&{ - 0.0001}\\
{ - 0.0006 - 0.0002i}&{0.0004 + 0.0002i}&{0.0003 - 0.0001i}&{0.0004 - 0.0005i}&{0.0039}&{0.0004 + 0.0021i}&{0.0050 - 0.0264i}&{0.00041 + 0.0001i}\\
{ - 0.0005 + 0.0005i}&{0.0004 + 0.0009i}&{0.0006 - 0.0007i}&{ - 0.0008 - 0.0010i}&{0.0004 - 0.0021i}&{0.0047}&{ - 0.0488 - 0.0249i}&{0.0002 - 0.0003i}\\
{0.0027 - 0.0024i}&{ - 0.0062 - 0.0068i}&{ - 0.0109 + 0.0198i}&{0.0219 + 0.0114i}&{0.0050 + 0.0264i}&{ - 0.0488 + 0.0249i}&{0.9818}&{0.001 + 0.0007i}\\
0&0&{0.0003}&{ - 0.0001}&{0.0004 - 0.0001i}&{0.0002 + 0.0003i}&{0.001 - 0.0007i}&{0.0027}
\end{array}} \right) \nonumber \\
\rho _{\exp }^{{\epsilon _2}} &= \left( {\begin{array}{*{20}{c}}
{0.0003}&{0.0002i}&{ - 0.0004}&{0.0004i}&{ - 0.0004 - 0.0003i}&{ - 0.0020 - 0.0014i}&{0.0051 - 0.0048i}&0\\
{0.0002i}&{0.0054}&{0.0012 - 0.0006i}&{ - 0.0002 - 0.0016i}&{ - 0.0009 + 0.0001i}&{0.0439 - 0.0030i}&{ - 0.0031 - 0.0084i}&0\\
{ - 0.0004}&{0.0012 + 0.0006i}&{0.0058}&{ - 0.0010 - 0.0020i}&{0.0006 + 0.0025i}&{0.0278 + 0.0094i}&{ - 0.0267 + 0.0019i}&0\\
{0.0004i}&{ - 0.0002 + 0.0016i}&{ - 0.0010 + 0.0020i}&{0.0026}&{ - 0.0025 + 0.0003i}&{ - 0.0113 + 0.0243i}&{ - 0.0098 - 0.0081i}&0\\
{ - 0.0004 + 0.0003i}&{ - 0.0009 - 0.0010i}&{0.0006 - 0.0025i}&{ - 0.0025 - 0.0003i}&{0.0031}&{0.0061 - 0.0214i}&{0.0062 + 0.0230i}&{0.0001i}\\
{ - 0.0012 + 0.0014i}&{0.0439 + 0.0030i}&{0.0278 - 0.0094i}&{ - 0.0113 - 0.0242i}&{0.0061 + 0.0214i}&{0.4981}&{ - 0.0189 - 0.0018i}&{0.0001i}\\
{0.0051 + 0.0048i}&{ - 0.0031 + 0.0084i}&{ - 0.0267 - 0.0019i}&{ - 0.0098 + 0.0081i}&{0.0062 - 0.0210i}&{ - 0.0189 + 0.0018i}&{0.4838}&0\\
0&0&0&0&{0.0001i}&{0.0001i}
\end{array}} \right)
\end{align}
\end{widetext}
\end{tiny}

The density matrices are shown graphically in Fig.~\ref{fig:matrices}.

\bibliography{topologicalbib}

\end{document}